\documentclass[english,aps,prc,nofootinbib,superscriptaddress,twocolumn,letter]{revtex4}
\usepackage[latin9]{inputenc}
\usepackage{hyperref}
\hypersetup{
  colorlinks=true,        % false: boxed links; true: colored links
  linkcolor=blue,         % color of internal links
  citecolor=cyan,         % color of links to bibliography
}
\usepackage{breakurl}
\usepackage{graphicx}
\usepackage{epsf}
\usepackage{epsfig}
\usepackage{amssymb,amsmath}
\usepackage[usenames]{color}
\usepackage{amssymb}
\usepackage{times}
\usepackage{comment}

\newcommand\snn{\sqrt{s_\text{NN}}}

\allowdisplaybreaks



\begin{document}

\preprint{This line only printed with preprint option}

%\title{Global and local polarization of $\Lambda$ hyperons in high energy collisions: effects of tilted bulk medium, velocity profile and its correlation with directed flow}

\title{Hyperon polarization and its relation with directed flow in high-energy nuclear collisions}

\author{Ze-Fang Jiang}
\email{jiangzf@mails.ccnu.edu.cn}
\affiliation{Department of Physics and Electronic-Information Engineering, Hubei Engineering University, Xiaogan, Hubei, 432000, China}
\affiliation{Institute of Particle Physics and Key Laboratory of Quark and Lepton Physics (MOE), Central China Normal University, Wuhan, Hubei, 430079, China}

\author{Xiang-Yu Wu}
\email{xiangyuwu@mails.ccnu.edu.cn}
\affiliation{Institute of Particle Physics and Key Laboratory of Quark and Lepton Physics (MOE), Central China Normal University, Wuhan, Hubei, 430079, China}

\author{Shanshan Cao}
\email{shanshan.cao@sdu.edu.cn}
\affiliation{Institute of Frontier and Interdisciplinary Science, Shandong University, Qingdao, Shandong, 266237, China}

\author{Ben-Wei Zhang}
\email{bwzhang@mail.ccnu.edu.cn}
\affiliation{Institute of Particle Physics and Key Laboratory of Quark and Lepton Physics (MOE), Central China Normal University, Wuhan, Hubei, 430079, China}
\affiliation{Guangdong Provincial Key Laboratory of Nuclear Science, Institute of Quantum Matter,
South China Normal University, Guangzhou, Guangdong, 510006, China}

\begin{abstract}
We investigate the hyperon polarization and its relation with the directed flow of the quark-gluon plasma (QGP) in non-central Au+Au collisions at $\snn=27$~GeV. A modified 3-dimensional (3D) Glauber model is developed and coupled to a (3+1)-D viscous hydrodynamic evolution of the QGP. Within this framework, we obtain a satisfactory simultaneous description of the directed flow of identified particles and $\Lambda$ polarization, and show sensitivity of polarization to both the tilted geometry and the longitudinal flow profile of the QGP. A non-monotonic transverse momentum dependence of the $\Lambda$ polarization is found in our calculation, which is absent from hydrodynamic simulation using other initialization methods and can be tested by future experimental data with higher precision. The relation between the global polarization and directed flow of $\Lambda$ is explored as the longitudinal flow field or the medium deformation varies. Due to  the common origin of these two observables, their combination may provide a more stringent constraint on the initial condition of the QGP.

\end{abstract}
%\pacs{20.24, 20.25}
\maketitle
\date{\today}

\section{Introduction}
\label{emsection1}

A highly excited state of nuclear matter, known as the strongly coupled quark-gluon plasma (QGP), is created in high-energy nucleus-nucleus collisions at the Relativistic Heavy-Ion Collider (RHIC) and the CERN Large Hadron Collider (LHC).
Since the discovery of the QGP at the beginning of this century, quantifying its properties becomes one of the primary goals of the heavy-ion collision programs~\cite{Ollitrault:1992bk,Rischke:1995ir,Sorge:1996pc,Bass:1998vz,Aguiar:2001ac,Shuryak:2003xe,Heinz:2013th,Busza:2018rrf}. 
In non-central heavy-ion collisions, huge orbital angular momentum (OAM) or vorticity field can be deposited into the QGP, leading to the global polarization of hyperons through the spin-orbital coupling~\cite{Liang:2004ph,Liang:2004xn,Gao:2007bc,Chen:2009sqr,Huang:2011ru} or spin-vorticity coupling~\cite{Becattini:2007sr,Becattini:2007nd,Becattini:2013fla,Becattini:2016gvu,Fang:2016vpj,Florkowski:2017dyn}. This initiates the exploration of spin physics in a strongly coupled system.
Chiral kinetic theory~\cite{Hidaka:2017auj,Shi:2020qrx,Fang:2022ttm}, spin-hydrodynamics~\cite{She:2021lhe,Xie:2023gbo,Biswas:2023qsw,Weickgenannt:2022qvh,Bhadury:2022ulr,Florkowski:2021wvk}, and their related phenomenologies, such as chiral vortical effect~\cite{Son:2009tf}, chiral vortical wave~\cite{Jiang:2015cva} and the change of the QCD phase diagram induced by the vorticity effect~\cite{Chen:2015hfc,Chernodub:2016kxh,Huang:2017pqe} are under active investigation.

Recently, the STAR experiment has confirmed the global polarization of $\Lambda(\bar{\Lambda})$ hyperons in semi-peripheral Au+Au collisions~\cite{STAR:2017ckg,STAR:2019erd,STAR:2023nvo}, which implies an average fluid vorticity of $\omega\approx (9\pm1) \times 10^{21}$ $s^{-1}$. This is the most vortical fluid ever observed in nature. Further analyses of the global and local polarization have revealed new insights into the vortical properties of the QGP~\cite{Gao:2020vbh,Huang:2020xyr}. 
Various theoretical approaches have been developed to study the influence of the fluid vorticity on spin polarization, including transport models (e.g. AMPT) with the assumption of local thermal equilibrium~\cite{Xia:2018tes,Wei:2018zfb,Guo:2019joy,Li:2017slc}, the quark-gluon-string model (QGSM)~\cite{Baznat:2017jfj} and (3+1)-dimensional viscous hydrodynamic models~\cite{Karpenko:2016jyx,Pang:2016igs,Fu:2020oxj,Ivanov:2019ern,Ivanov:2020udj,Wu:2019eyi,Ryu:2021lnx,Yi:2021ryh,Alzhrani:2022dpi,Wu:2022mkr}. 
These models consistently capture the features of the beam energy dependence of the global polarization along the out-of-plane direction ($-P^{y}$) as observed from the RHIC to the LHC energies. However, inconsistency still remains in the azimuthal angle dependence of the local polarization between theoretical calculations and experimental data~\cite{Niida:2018hfw,STAR:2019erd}. Considerable efforts have been devoted in resolving this local polarization puzzle~\cite{Sun:2018bjl,Yang:2018lew,Liu:2020dxg,Becattini:2021iol,Fu:2021pok,Liu:2021uhn,Guo:2021udq,She:2021lhe,Fu:2022myl}.

Within the hydrodynamic approach, it has been found that the hyperon polarization is sensitive to the initial condition of the QGP evolution~\cite{Betz:2007kg,Alzhrani:2022dpi,Wu:2022mkr,Jiang:2023fad}. Significant impacts on the polarization have been revealed from the initial velocity field of the medium~\cite{Alzhrani:2022dpi,Wu:2022mkr,Wu:2019eyi}, and the initial geometry of the medium which affects the vorticity field inside the QGP~\cite{Voloshin:2017kqp}. Since these aspects are also the origin of other soft hadron observables like their collective flow coefficients, it would be of great interest to study polarization together with these observables in the same framework and utilize their combination to better constrain the initial condition of the QGP~\cite{ALICE:2013xri,STAR:2016cio,ALICE:2022wpn,STAR:2014clz}. This is the focus of our present work.

Following our previous exploration ~\cite{Jiang:2021foj,Jiang:2023fad} on the interplay between the hydrodynamic initial condition and the directed flow of hadrons in non-central heavy-ion collisions, we will further investigate how the tilted geometry of the QGP fireball and its longitudinal flow velocity field affect the hyperon polarization, including its dependence on rapidity, centrality and transverse momentum. Detailed comparisons on the $\Lambda$ polarization will be conducted between different contributions to polarization from the kinetic theory, and also between different initialization models of our hydrodynamic simulation. Since the asymmetric initial condition serves as the common origin of both the hyperon polarization and the directed flow of hadrons, we will explore the relation between these two observables as the medium geometry and the flow field vary. We will use  Au+Au collisions at $\snn=27$~GeV as an environment for our discussion, considering the abundance of experimental data on both polarization and directed flow coefficient of $\Lambda$ hyperons in this collision system.

The rest of this paper will be structured as follows. In Sec.~\ref{v1section2}, we will first present the theoretical framework we develop for a simultaneous investigation on directed flow and polarization of the QGP, including a 3-dimensional (3D) Glauber model that involves a tilted medium geometry and an initial longitudinal flow field, a (3+1)-D hydrodynamic model for the QGP evolution and a modified Cooper-Frye formalism for evaluating the polarization pseudo-vector on the chemical freezeout hypersurface. Numerical results on the hadron directed flow and the hyperon polarization will then be presented in Sec.~\ref{v1section3}, with specific focus on the dependence of the $\Lambda$ polarization on the medium geometry and longitudinal flow profile, and the relation between polarization and directed flow. In the end, we will summarize in Sec.~\ref{v1section4}.

\section{Model framework}
\label{v1section2}

\subsection{Initial condition}
\label{v1subsect2}

%%%  Fig-1 
\begin{figure}[tbp!]
\begin{center}
\includegraphics[width=0.75\linewidth]{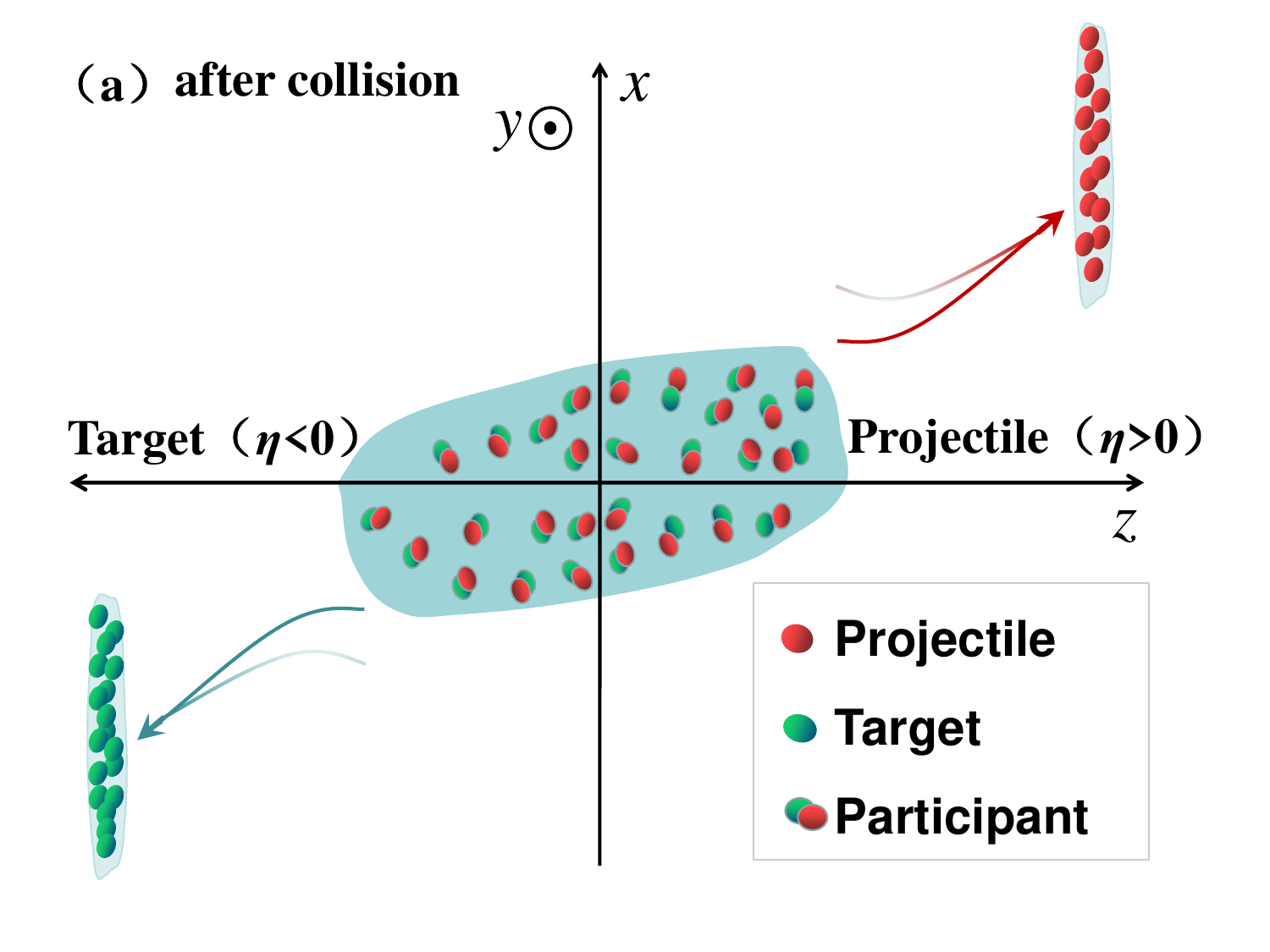}
\includegraphics[width=0.80\linewidth]{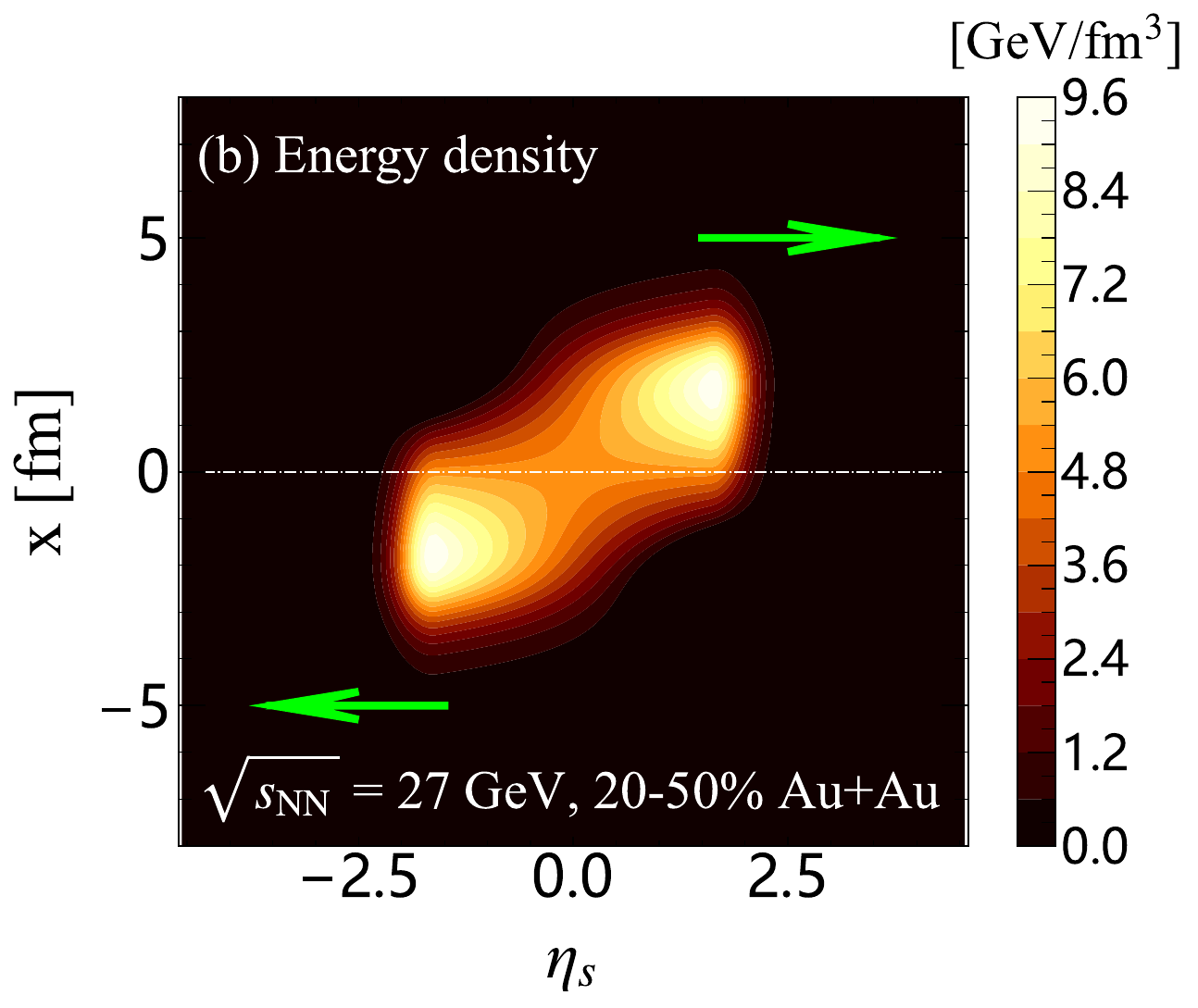}
\includegraphics[width=0.80\linewidth]{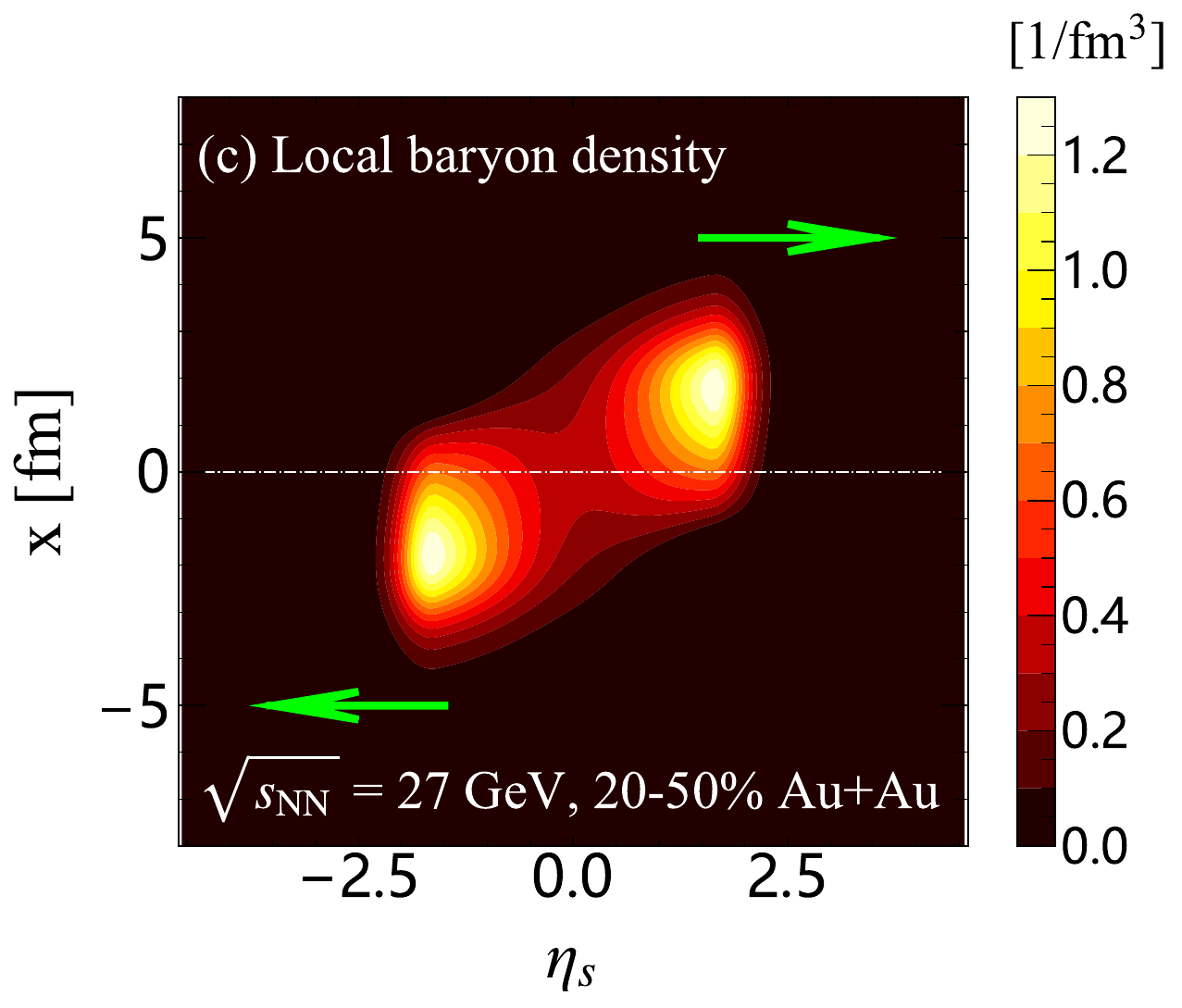}
\end{center}
\caption{(Color online) 
Upper panel: illustration of non-central heavy-ion collisions in the reaction plane, where two nuclei travel along $\pm z$ direction, collide at $z=0$, and deposit global orbital angular momentum (OAM) along the $-y$ direction. 
Middle and lower panels: 
the initial energy density and net baryon number density profiles, respectively, on the $\eta_\text{s}$-$x$ plane for 20-50\% Au+Au collisions at $\sqrt{s_\text{NN}}=27$~GeV, with arrows denoting propagation towards forward and backward rapidities.}
\label{f:ed}
\end{figure}

We use a modified Glauber model to generate the initial condition of hydrodynamic evolution of the QGP, which possesses a counterclockwisely tilted geometry in the reaction plane with respect to the beam (longitudinal) direction~\cite{Bozek:2010bi,Jiang:2021foj,Jiang:2021ajc}.

The Woods-Saxon (WS) distribution of nucleons is applied to calculate the nuclear thickness function of the Au nucleus as
\begin{equation}
\begin{aligned}
T(x,y)=\int_{-\infty}^{\infty}dz\frac{\rho_{0}}{1+\exp\left[(r-R_{0})/d_{0}\right]},
\label{eq:thicknessf}
\end{aligned}
\end{equation}
where $\rho_0=0.17~$fm$^{-3}$ is the average nucleon density, $r=\sqrt{x^{2}+y^{2}+z^{2}}$ is the radial position with $x,~y,~z$ being the space coordinates, $R_{0}=6.38$ fm is the radius of nucleus and $d_{0}=0.535$~fm is the surface diffusiveness parameter.

For two nuclei travelling along the longitudinal ($\pm \hat{z}$) direction and colliding with an impact parameter $\mathbf{b}$, their thickness functions are then given by
\begin{equation}
\begin{aligned}
T_{+}(\mathbf{x}_\text{T})=T(\mathbf{x}_\text{T}-\mathbf{b}/2),~~~~T_{-}(\mathbf{x}_\text{T})=T(\mathbf{x}_\text{T}+\mathbf{b}/2),
\label{eq:t+}
\end{aligned}
\end{equation}
where $\mathbf{x}_\text{T}=(x,y)$ is the transverse plane coordinate. According to the Glauber model, their corresponding densities of participant nucleons of inelasitic scatterings are given by 
\begin{align}
T_{1}(\mathbf{x}_\text{T})&=T_{+}(\mathbf{x}_\text{T})\left\{1-\left[1-\frac{\sigma_\text{NN} T_{-}(\mathbf{x}_\text{T})}{A}\right]^{A}\right\}  \, ,\\
T_{2}(\mathbf{x}_\text{T})&=T_{-}(\mathbf{x}_\text{T})\left\{1-\left[1-\frac{\sigma_\text{NN} T_{+}(\mathbf{x}_\text{T})}{A}\right]^{A}\right\}  \, ,
\end{align}
with $A$ being the mass number and $\sigma_\text{NN}$ being the inelastic nucleon-nucleon scattering cross section~\cite{Loizides:2017ack}.

Inspired by the anisotropy of hadrons emitted by the QGP, it has been proposed in Ref.~\cite{Bozek:2011ua} that non-central collisions deposit energy asymmetrically along the longitudinal direction, as illustrated in the upper panel of Fig.~\ref{f:ed}. This leads to a counterclockwise tilt of the QGP fireball in the reaction plane with respect to the beam direction. Different parameterization schemes of the initial condition have been proposed in literature~\cite{Bozek:2016kpf,Jiang:2021foj,Jiang:2021ajc} to introduce this deformation of the nuclear matter and have been shown consistent with each other. In this work, we follow our earlier studies~\cite{Jiang:2021foj,Jiang:2021ajc} and parameterize the spacetime rapidity ($\eta_\text{s}$) dependence of wounded (or participant) nucleon distribution as
\begin{equation}
\begin{aligned}
W_\text{N}(x,y,\eta_\text{s})=&~T_{1}(x,y)+T_{2}(x,y) \\
+&~H_\text{t}[T_{1}(x,y)-T_{2}(x,y)]\tan\left(\frac{\eta_\text{s}}{\eta_\text{t}}\right),
\label{eq:mnccnu}
\end{aligned}
\end{equation}
where the parameter $H_\text{t}$ reflects the overall imbalance strength of energy deposition between forward and backward $\eta_\text{s}$. It relies on the impact parameter of collisions, and is set as $H_\text{t} = 2.07b/\textrm{fm}$ in the present study in order to consistently describe the centrality dependence of the soft hadron observables in Au+Au collisions at $\snn=27$~GeV later. Additionally, the function $\tan (\eta_\text{s}/\eta_\text{t})$ in Eq.~(\ref{eq:mnccnu}) determines how the imbalance varies with $\eta_\text{s}$. 
We use a constant parameter $\eta_\text{t}=8.0$ in this study, which provides a good description of the directed flow ($v_1$) of charged particles in our previous work~\cite{Jiang:2021foj}.

After accounting for contributions from both wounded nucleons and binary (hard) collisions, the total weight function reads
\begin{equation}
\begin{aligned}
W(x,y,\eta_\text{s})=\frac{(1-\alpha)W_\text{N}(x,y,\eta_\text{s})+\alpha n_\text{BC}(x,y)}{\left[(1-\alpha)W_\text{N}(0,0,0)+\alpha n_\text{BC}(0,0)\right]|_{\mathbf{b}=0}},
\label{eq:wneta}
\end{aligned}
\end{equation}
where $n_\text{BC}(x,y)=\sigma_\text{NN}T_{+}(x,y)T_{-}(x,y)$ is the number of binary collisions, and $\alpha=0.05$ is called the collision hardness parameter determined by the centrality (or $\mathbf{b}$) dependence of the soft hadron yield~\cite{Pang:2018zzo,Loizides:2017ack}.

Under the Bjorken flow assumption, the initial energy density $\varepsilon_0$ and the normalized local net baryon density $n_0$ are given by~\cite{Wu:2021fjf,Pang:2018zzo}
\begin{align}
\label{eq:eps} \varepsilon_{0}(x,y,\eta_\text{s})&=K \cdot W(x,y,\eta_\text{s}) \cdot H(\eta_\text{s}) \, ,\\
\label{eq:nb} n_{0}(x,y,\eta_\text{s})&=\frac{1}{N}\cdot W(x,y,\eta_\text{s}) \cdot H(\eta_\text{s}) \cdot H_{B}(\eta_\text{s})  \, ,
\end{align}
with the overall factor $K$ set by the multiplicity distribution ($dN_{\textrm{ch}}/d\eta$ or $dN_{\textrm{ch}}/dy$) of soft hadrons, and $N$ being a normalization factor for $n_0$.

In Eqs.~(\ref{eq:eps}) and~(\ref{eq:nb}), a function
\begin{equation}
\begin{aligned}
H(\eta_\text{s})=\exp\left[-\frac{(|\eta_\text{s}|-\eta_\text{w})^{2}}{2\sigma^{2}_{\eta}}\theta(|\eta_\text{s}|-\eta_\text{w}) \right]
\label{eq:heta}
\end{aligned}
\end{equation}
is introduced to describe the plateau structure in the longitudinal distribution of emitted hadrons, in which $\eta_\text{w}$ controls the width of the central rapidity plateau and $\sigma_{\eta}$ determines the width (speed) of the Gaussian decay outside the plateau region~\cite{Pang:2018zzo}. In order to model the accumulation of baryons in the forward and backward rapidity regions, we also include the following distribution of baryon density in the longitudinal direction~\cite{Bozek:2022svy,Jiang:2023fad} 
\begin{equation}
\begin{aligned}
H_{B}(\eta_\text{s})=\exp\left[-\frac{(\eta_\text{s}-\eta_{n})^{2}}{2\sigma^{2}_{n}}\right]+\exp\left[-\frac{(\eta_\text{s}+\eta_{n})^{2}}{2\sigma^{2}_{n}}\right],
\label{eq:hetan}
\end{aligned}
\end{equation}
where parameters $\eta_{n}$ and $\sigma_{n}$ are calibrated by the $p_\text{T}$ spectra of protons and antiprotons~\cite{Jiang:2023fad}. 

Since we aim at exploring the hyperon polarization in the same framework, which is sensitive to the gradient of the fluid velocity field~\cite{Li:2022pyw}, we need to extend the initialization model beyond the Bjorken approximation for the fluid velocity. Following Refs.~\cite{Shen:2020jwv, Ryu:2021lnx, Alzhrani:2022dpi}, we construct the initial energy-momentum tensor components as
\begin{align}
\label{eq:Ttautau}
T^{\tau\tau}&=\varepsilon_{0}(x,y,\eta_\text{s})\cosh(y_\text{L}) \, ,\\
\label{eq:Ttaueta}
T^{\tau\eta_\text{s}}&=\frac{1}{\tau_{0}}\varepsilon_{0}(x,y,\eta_\text{s})\sinh(y_\text{L})  \, ,
\end{align}
where the rapidity variable is modeled as
\begin{equation}
\begin{aligned}
y_\text{L} \equiv f_{v} y_{\textrm{CM}}.
\label{eq:yl}
\end{aligned}
\end{equation}
Here, the center of mass rapidity $y_{\textrm{CM}}$ at a given transverse location $(x,y)$ depends on both the beam energy  $y_{\textrm{beam}}\equiv\textrm{arccosh}[\sqrt{s_{\textrm{NN}}}/(2m_{\textrm{N}})]$ and the imbalance between the participant thickness functions as
\begin{equation}
\begin{aligned}
y_{\textrm{CM}}=\textrm{arctanh} \left[\frac{T_{1}-T_{2}}{T_{1}+T_{2}} \tanh (y_{\textrm{beam}})\right],
\label{eq:ycm}
\end{aligned}
\end{equation}
where $m_{\textrm{N}}$ is the nucleon mass and $f_{v} \in [0, 1]$ parameterizes the fraction of $y_{\textrm{CM}}$ deposited into the longitudinal flow velocity.
This $f_{v}$ parameter allows one to vary the magnitude of the longitudinal flow velocity gradient, which influences both local and global polarization of $\Lambda(\bar{\Lambda})$ hyperons. When $f_{v}=0$, one recovers the Bjorken flow scenario with $y_\text{L}=0$~\cite{Shen:2020jwv}. With Eqs.~(\ref{eq:Ttautau}) and~(\ref{eq:Ttaueta}), the initial fluid velocity in the $\eta_\text{s}$ direction is given by
$v_{\eta_\text{s}}=T^{\tau\eta_\text{s}}/(T^{\tau\tau}+P)$, in which $P$ is the pressure.
In the present work, the initial fluid velocity in the transverse plane is assumed to be zero by setting $T^{\tau x} = T^{\tau y} = 0$. 
%{\color{red} The local energy density and  baryon density with finite initial flow velocity assumption can be found according to the root finding algorithm for initial energy-momentum tensor and initial baryon current~\cite{Ryu:2021lnx}.} 

\begin{table}[h]
\centering
\vline
\begin{tabular}{c|c|c|c|c|c|c|c|c|}
\hline
$\sqrt{s_\text{NN}}$ [GeV]& $K$ & $\tau_0$ [fm] & $\sigma_\eta$ [fm] & $\eta_\text{w}$ & $\sigma_{n}$ &$\eta_{n}$ & $f_v$  \\ \hline
27   & 7.40  & 1.4 & 0.3 & 1.6   & 1.06  &1.8  &0.23  \\ \hline
% 7.7  & 2.50  & 2.6 & 0.3 & 0.9   & 0.70  &1.05 &0.26  \\ \hline
\end{tabular}
\caption{Parameters of the initial condition based on the 3-dimensional tilted optical Glauber model~\cite{Wu:2021fjf,Shen:2020jwv,Jiang:2023fad}.}
\label{table:parameters}
\end{table}

In Tab.~\ref{table:parameters}, we summarize the parameters used to initialize the QGP medium in this study. The first four parameters ($K$, $\tau_0$, $\sigma_\eta$, and $\eta_\text{w}$) are adjusted based on the rapidity dependence of the charged particle yields ($dN_\text{ch}/dy$) in the most central collisions at a given beam energy. 
With these parameters, the combination of our initial condition and the CLVisc hydrodynamic simulation is able to provide a good description of the $p_\text{T}$ spectra of different types of identified particles ($\pi^+$, $K^+$, $p$ and $\bar{p}$) in different centrality regions across the RHIC-BES energies~\cite{Jiang:2023fad}. This provides a reliable baseline for our subsequent investigation on the global and local polarization of hyperons in this work. 
%The initial velocity and deformed geometry of the QGP medium are determined by the parameters $f_{v}$ and $H_\text{t}$, which have little effect on the particle spectra integrated over the azimuthal angle.~\cite{Bozek:2011ua,Jiang:2021ajc,Jiang:2021foj,Bozek:2022svy,Wu:2021fjf,Shen:2020jwv,Ryu:2021lnx}. 
The last parameter ($f_{v}$) in In Tab.~\ref{table:parameters} is adjusted according to the directed flow coefficients of $\pi^{-}$, $p$ and $\bar{p}$, $\Lambda$ and $\bar{\Lambda}$. The value of $f_v$ we use here is different from the one used in Ref.~\cite{Ryu:2021lnx} due to our different assumptions on the initial geometry of the QGP profile. With the decrease of the beam energy, a larger fraction of the longitudinal momentum of the colliding nuclei can be deposited into the initial longitudinal velocity~\cite{Jiang:2023fad}.  

With the model parameters listed above, we first present in Fig.~\ref{f:ed} the distributions of the initial energy density (middle panel) and net baryon number density (bottom panel) on the $\eta_\text{s}$-$x$ plane for 20-50\% ($b=8.57$~fm) Au+Au collisions at $\snn=27$~GeV. Their values beyond the Bjorken approximation are solved from the modified energy-momentum tensor components in Eqs.~(\ref{eq:Ttautau}) and~(\ref{eq:Ttaueta}). One may clearly observe a tilted geometry of the QGP fireball with respect to the beam direction within this initialization model. Apart from an asymmetrical shift along the forward and backward rapidity directions, a counterclockwise tilt in the $\eta_\text{s}$-$x$ plane can be seen for both the energy and net baryon densities. Due to their different parameterizations in Eq.~(\ref{eq:eps}) and Eq.~(\ref{eq:nb}), the baryon density exhibits stronger shift towards large rapidity as well as stronger tilt compared to the energy density. As discussed in Ref.~\cite{Bozek:2022svy}, this could be understood with the string models of the initial state~\cite{Shen:2017bsr,Bialas:2004kt,Jezabek:2021oxg}: while the baryon density deposition is driven by the valence quarks in the participant nucleons, energy density deposition originates from the melting of strings that involves both valence and sea quarks. We expect stronger tilt of these density profiles in more peripheral collisions due to the stronger drag experienced by participant nucleons from spectators. In phenomenology, the asymmetry in the energy density is responsible for the rapidity-odd directed flow of soft hadrons, while the asymmetry in the baryon density affects the abundance of baryons and anti-baryons produced from different locations of the QGP~\cite{Bozek:2022svy}.

%We find that the baryon density profile is more tilted at peripheral collisions because the participant nucleons experience a stronger drag from spectators at lower collisional energies.
%Meanwhile, the colliding beams deposit more fractional longitudinal momentum into the QGP at lower energies, as reflected by the increasing value of $f_v$ as $\snn$ decreases. 
%We also there is a stronger tilt of the baryon density than the energy density by using $1.2 H_\text{t}$ for the former at very low energy (for 7.7~GeV), to better describe the proton $v_1$ phenomenologically~\cite{Jiang:2023fad}. 
%This might result from effects of the phase transition~\cite{Konchakovski:2014gda,Ivanov:2014ioa,Steinheimer:2014pfa,Nara:2016phs}, emission of the spectator matter~\cite{Zhang:2018wlk} and the strong electromagnetic field~\cite{Rybicki:2013qla} that have not considered in our current work.

\subsection{Hydrodynamic evolution}

Starting with the initial condition constructed in the previous subsection, we utilize a (3+1)-D viscous hydrodynamic model CLVisc~\cite{Pang:2016igs,Pang:2018zzo,Wu:2018cpc,Wu:2021fjf} to describe the further evolution of the QGP medium. Under finite baryon chemical potential, the hydrodynamic equations read~\cite{Jiang:2020big,Jiang:2018qxd,Denicol:2012cn,Romatschke:2009im,Romatschke:2017ejr}
\begin{align}
\nabla_{\mu} T^{\mu\nu}&=0 \, ,\\
\nabla_{\mu} J^{\mu}&=0  \, ,
\end{align}
where the energy-momentum tensor $T^{\mu\nu}$ and the net baryon current $J^{\mu}$ are defined as
\begin{align}
T^{\mu\nu} &= \varepsilon U^{\mu}U^{\nu} - P\Delta^{\mu\nu} + \pi^{\mu\nu}\,, \\	
J^{\mu} &= nU^{\mu}+V^{\mu}\,,
\end{align}
with $\varepsilon$, $P$, $n$, $u^{\mu}$, $\pi^{\mu\nu}$, $V^{\mu}$ being the local energy density, pressure, net baryon density, flow velocity field, shear stress tensor and baryon diffusion current respectively.
The projection tensor is given by $\Delta^{\mu\nu} = g^{\mu\nu}-u^{\mu}u^{\nu}$ with the metric tensor $g^{\mu\nu} = \text{diag} (1,-1,-1,-1)$. Effects of the bulk viscosity are not included in the present study yet.

The dissipative currents $\pi^{\mu\nu}$ and $V^{\mu}$ are given by the following expressions based on the Israel-Stewart-like second order hydrodynamic expansion~\citep{Denicol:2018wdp}:

\begin{align}
\Delta^{\mu\nu}_{\alpha\beta} (u\cdot \partial) \pi^{\alpha\beta} = &
 -\frac{1}{\tau_{\pi}}\left(\pi^{\mu\nu} - \eta_\text{v}\sigma^{\mu\nu}\right)
- \frac{4}{3}\pi^{\mu\nu}\theta
\nonumber
\\
&
-\frac{5}{7}\pi^{\alpha<\mu}\sigma_{\alpha}^{\nu>}+ \frac{9}{70}\frac{4}{e+P}\pi^{<\mu}_{\alpha}\pi^{\nu>\alpha}\,,
\nonumber
\\
\Delta^{\mu\nu} (u\cdot \partial) V_{\nu}  = &  - \frac{1}{\tau_V}\left(V^{\mu}-\kappa_B\bigtriangledown^{\mu}\frac{\mu_B}{T}\right)-V^{\mu}\theta
\nonumber \\
&-\frac{3}{10}V_{\nu}\sigma^{\mu\nu}\,,
\end{align}
where $\theta = \partial \cdot u$ is the expansion rate, $\sigma^{\mu\nu} = \partial^{<\mu} u^{\nu>}$ is the shear tensor,
$\eta_\text{v}$ and $\kappa_B$ are the shear viscosity and baryon diffusion coefficient.
For an arbitrary tensor $A^{\mu\nu}$, its traceless symmetric part is given by $A^{<\mu\nu>} = \frac{1}{2}[(\Delta^{\mu\alpha}\Delta^{\nu\beta}+\Delta^{\nu\alpha}\Delta^{\mu\beta})-\frac{2}{3}\Delta^{\mu\nu}\Delta^{\alpha\beta}]A_{\alpha \beta}$~\cite{Wu:2021fjf}.

The specific shear viscosity $C_{\eta_\text{v}}$ and the baryon diffusion coefficient $\kappa_B$ are model parameters in hydrodynamic simulation, which are connected to $\eta_\text{v}$ and parameter $C_B$ via
\begin{align}
C_{\eta_\text{v}} &= \frac{\eta_\text{v} T}{e+P}, \label{eq:C_shear}\\
\kappa_B &= \frac{C_B}{T}n\left[\frac{1}{3} \cot \left(\frac{\mu_B}{T}\right)-\frac{nT}{e+P}\right] \,, \label{eq:CB}
\end{align}
where $\mu_B$ is the baryon chemical potential. In this work, we use $C_{\eta_\text{v}}=0.08$ and ${C_B}=0.4$ for all collision centrality classes~\cite{Wu:2021fjf}, and set the relaxation times as $\tau_{\pi} = 5C_{\eta_\text{v}}/T$ and $\tau_V = C_B/T$.

We solve the hydrodynamic equations using the NEOS-BQS equation of state (EOS)~\cite{Monnai:2019hkn,Monnai:2021kgu}, which extends the lattice EOS at zero net baryon density to finite net baryon density via the Taylor expansion~\cite{Monnai:2019hkn,Monnai:2021kgu}. This EOS provides a smooth crossover between the QGP and the hadron phase under the conditions of strangeness neutrality ($n_S=0$) and electric charge density $n_Q = 0.4n_B$.

\subsection{Particlization}

We use the isoenergy-density freezeout condition~\cite{Pang:2018zzo} in our study and determine the freezeout hypersurface by a fixed energy density ($e_{\text{frz}}$= 0.4~GeV/fm$^3$)~\cite{Wu:2021fjf}. We apply the Cooper-Frye formalism on this hypersurface to obtain the hadron momentum distribution:
\begin{align}
\frac{dN}{p_\text{T} dp_\text{T} d\phi dy } = \frac{g_i}{(2\pi)^3}\int_{\Sigma} p^{\mu}d\Sigma_{\mu}f_\mathrm{eq}(1+\delta f_{\pi}+\delta f_{V})\,.
\end{align}
In the above equation, $g_i$ is the spin-color degeneracy factor for identified hadrons, and $d\Sigma_{\mu}$ is the hypersurface element determined by the projection method~\cite{Pang:2018zzo}. The thermal distribution ($f_{\rm eq}$) and the out-of-equilibrium corrections ($\delta f_{\pi}$ and $\delta f_{V}$) satisfy
\begin{align}
	&f_{\rm eq} = \frac{1}{\exp \left[(p_{\mu}U^{\mu} - B\mu_B \right)/T_\text{f}] \mp 1} \, ,\\
	&\delta f_{\pi} = (1\pm f_{\text{eq}}) \frac{p_{\mu}p_{\nu}\pi^{\mu\nu}}{2T^2_\text{f}(e+P)}, \\
	&\delta f_V = (1\pm f_{\text{eq}})\left(\frac{n_B}{e+P}-\frac{B}{U^{\mu}p_{\mu}}\right)\frac{p^{\mu}V_{\mu}}{\kappa_B/ \tau_V },
\end{align}
where $T_\text{f}$ is the chemical freezeout temperature, and $B$ represents the baryon number of an identified hadron. 
The out-of-equilibrium corrections above are obtained from the Boltzmann equation via the relaxation time approximation~\citep{McNelis:2021acu}. 
We take into account the resonance decay contributions, as conducted in Ref.~\cite{Pang:2018zzo}, when evaluating the directed flows of pions, protons and $\Lambda$ hyperons. However, the $\Lambda$ polarization is still computed directly on the hadronization hypersurface, as discussed in the next subsection, without contributions from these resonance decays. Effects of hadronic scatterings after the QGP phase have not been included in the present work.
%{\color{blue} In our current study, we compute the polarization of only primary $\Lambda$ hyperons that directly emit from the fluid (i.e., $e_{\text{frz}}$= 0.4~GeV/fm$^3$ hypersurface). 
%The $\Lambda$ hyperons arising from the decays of heavier hyperons(e.g. $\Sigma^{0}\rightarrow$ $\Lambda+\gamma$, $\Sigma^{*}$ $\rightarrow$ $\Lambda+\pi$, $\Xi$ $\rightarrow$ $\Lambda+\pi$) are not included in our analysis. We have taken into account the resonance decay contributions to calculate the directed flow of final pion, proton, and $\Lambda$ hyperons in this work~\cite{Pang:2018zzo}. 
%However, hadronic scatterings after the QGP phase have not been included yet.}

%% Section - II D
\subsection{Spin polarization}

In non-central heavy-ion collisions, the quarks are polarized due to the massive initial orbital angular momentum of the QGP fireball~\cite{Liang:2004ph,Betz:2007kg}. 
We assume the collision system to be in local thermal equilibrium on the freezeout hypersurface. 
%{\color{blue} Meanwhile, the conservation of spin is respected during hadronization and resonance decay processes.} 
The polarization pseudo-vector for spin-1/2 fermions can be obtained using the modified Cooper-Frye formalism as~\citep{Becattini:2013fla,Fang:2016vpj}
\begin{equation}
\mathcal{S}^{\mu}(\mathbf{p})=\frac{\int d \Sigma \cdot p \mathcal{J}_{5}^{\mu}(p, X)}{2 m \int d \Sigma \cdot \mathcal{N}(p, X)},
\end{equation}
where $\mathcal{J}^{\mu}_5$ is the axial charge current density and $\mathcal{N}^{\mu}(p, X)$ is the number density of fermions in the phase space.
Based on the decomposition of the vector product between the thermal vorticity tensor and the 4-momentum vector~\citep{Karpenko:2018erl} or the quantum kinetic theory~\citep{Yi:2021ryh,Hidaka:2017auj,Yi:2021unq}, 
%{\color{blue}Based on the result of the decomposition of the vector product of the thermal vorticity tensor and 4-momentum vector~\citep{Karpenko:2018erl} or the quantum kinetic theory ~\citep{Yi:2021ryh,Hidaka:2017auj,Yi:2021unq},}
$\mathcal{S}^{\mu}(\mathbf{p})$ can be decomposed into different sources as
\begin{eqnarray}
\mathcal{S}^{\mu}(\mathbf{p}) & = & \mathcal{S}_{\textrm{thermal}}^{\mu}(\mathbf{p})
+\mathcal{S}_{\textrm{shear}}^{\mu}(\mathbf{p})+\mathcal{S}_{\textrm{accT}}^{\mu}(\mathbf{p})   \nonumber \\
& &+\mathcal{S}_{\textrm{chemical}}^{\mu}(\mathbf{p})+\mathcal{S}_{\textrm{EB}}^{\mu}(\mathbf{p}),
\label{eq:totS}
\end{eqnarray}
where
\begin{eqnarray}
\mathcal{S}_{\textrm{thermal}}^{\mu}(\mathbf{p}) & = & \int d\Sigma^{\sigma}F_{\sigma}\epsilon^{\mu\nu\alpha\beta}p_{\nu}\partial_{\alpha}\frac{u_{\beta}}{T},\nonumber \\
\mathcal{S}_{\textrm{shear}}^{\mu}(\mathbf{p}) & = & \int d\Sigma^{\sigma}F_{\sigma}  \frac{\epsilon^{\mu\nu\alpha\beta}p_{\nu} u_{\beta}}{(u\cdot p)T}
 \nonumber \\
 & &\times  p^{\rho}(\partial_{\rho}u_{\alpha}+\partial_{\alpha}u_{\rho}-u_{\rho}Du_{\alpha}), \nonumber \\
\mathcal{S}_{\textrm{accT}}^{\mu}(\mathbf{p}) & = & -\int d\Sigma^{\sigma}F_{\sigma}\frac{\epsilon^{\mu\nu\alpha\beta}p_{\nu}u_{\alpha}}{T}
\left(Du_{\beta}-\frac{\partial_{\beta}T}{T}\right),\nonumber \\
\mathcal{S}_{\textrm{chemical}}^{\mu}(\mathbf{p}) & = & 2\int d\Sigma^{\sigma}F_{\sigma}\frac{1}{(u\cdot p)}\epsilon^{\mu\nu\alpha\beta}p_{\alpha}u_{\beta}\partial_{\nu}\frac{\mu}{T},\nonumber \\
\mathcal{S}_{\textrm{EB}}^{\mu}(\mathbf{p}) & = & 2\int d\Sigma^{\sigma}F_{\sigma}\left[\frac{\epsilon^{\mu\nu\alpha\beta}p_{\alpha}u_{\beta}E_{\nu}}{(u\cdot p)T}+\frac{B^{\mu}}{T}\right],\nonumber \label{eq:S_all}  \\
\label{eq:5S}
\end{eqnarray}
with
\begin{align}
&F^{\mu} = \frac{\hbar}{8m_{\Lambda}\Phi(\mathbf{p})}p^{\mu}f_\text{eq}(1-f_\text{eq}), \nonumber \\
&\Phi(\mathbf{p}) = \int d\Sigma^{\mu}p_{\mu}f_\text{eq}.
\label{eq:def_N}
\end{align}
The five terms in Eq.~(\ref{eq:5S}) represent polarization induced by the thermal vorticity ($\mathcal{S}_{\textrm{thermal}}^{\mu}$), the shear tensor ($\mathcal{S}_{\textrm{shear}}^{\mu}$),
the fluid acceleration minus temperature gradient ($\mathcal{S}_{\textrm{accT}}^{\mu}$),
the gradient of chemical potential over temperature ($\mathcal{S}_{\textrm{chemical}}^{\mu}$),
and the external electromagnetic field ($\mathcal{S}_{\textrm{EB}}^{\mu}$), respectively.
Detailed expressions of these terms can be derived from the gradient expansion method at local equilibrium~\cite{Becattini:2019dxo,Becattini:2021suc,Becattini:2021iol} or the Kubo formula~\cite{Liu:2020dxg,Liu:2021uhn,Fu:2021pok,Fu:2022myl}.
Here, $S^\mu_{\textrm{shear}}$ and $S^\mu_{\textrm{chemical}}$ are also named as the shear-induced polarization (SIP) and the baryonic spin Hall effect (SHE) in literature~\cite{Liu:2020dxg}.
Since the electromagnetic field decays rapidly and its evolution profile has not been well constrained in heavy-ion collisions yet, we only take the first four terms but neglect $S^{\mu}_\mathrm{EB}$.

The polarization vector of $\Lambda$ (or $\bar{\Lambda}$) in its rest frame can then be constructed as
\begin{eqnarray}
\vec{P}^{*}(\mathbf{p}) = \vec{P}(\mathbf{p})-\frac{\vec{P}(\mathbf{p}) \cdot \mathbf{p}}{p^{0}(p^{0}+m)}\mathbf{p},
\end{eqnarray}
where 
\begin{eqnarray}
P^{\mu}(\mathbf{p}) \equiv \frac{1}{s} \mathcal{S}^{\mu}(\mathbf{p}),
\end{eqnarray}
with $s=1/2$ being the particle spin.
After averaging over the transverse momentum, one obtains the local polarization as
\begin{eqnarray}
\langle \vec{P}(\phi_p) \rangle = \frac{\int_{y_{\text{min}}}^{y_{\text{max}}}dy \int_{p_\text{T\text{min}}}^{p_\text{T\text{max}}}p_\text{T}dp_\text{T}
[ \Phi (\mathbf{p})\vec{P}^{*}(\mathbf{p})]}{\int_{y_{\text{min}}}^{y_{\text{max}}}dy \int_{p_\text{T\text{min}}}^{p_\text{T\text{max}}}p_\text{T}dp_\text{T} \Phi(\mathbf{p}) },
\label{eq:localP}
\end{eqnarray}
in which $\phi_p$ is the azimuthal angle, and $\Phi(\mathbf{p})$ is an integration on the freezeout hypersurface defined in Eq.~(\ref{eq:def_N}). 
The mass of $\Lambda$ (or $\overline{\Lambda}$) is set as $m = 1.116$~GeV. Finally, the global polarization of $\Lambda$ and $\overline{\Lambda}$ is obtained by further averaging $\vec{P}^{*}(\mathbf{p})$ over $\phi_p$ in Eq.~(\ref{eq:localP}). 
Contributions to the $\Lambda$ polarization from resonance decays have not been included in this work, which could be improved according to Ref.~\cite{Becattini:2016gvu} in our future effort.
% {\color{blue} Contributions to the $\Lambda$ polarization from resonance decays have not been included in this work, which could be improved according to Ref.~\cite{Becattini:2016gvu} in our future effort.}

%%%%%%  Section 3 
\section{Numerical results}
\label{v1section3}

In this section, we present the directed flow coefficient and polarization of $\Lambda$($\bar{\Lambda}$) hyperons in Au+Au collisions at $\snn=$ 27~GeV from the CLVisc hydrodynamic calculation using the tilted initial geometry with non-zero initial longitudinal flow velocity field.
We first analyze the directed flow $v_{1}$ of pions, protons and antiprotons in various centrality classes to determine the $H_\text{t}$ value for the tilted QGP fireballs at different centralities. Using the $H_\text{t}$ value extracted from the directed flow, we then investigate the relation between the global polarization of $\Lambda(\bar{\Lambda})$ hyperons and centrality, transverse momentum, and pseudo-rapidity in Sec.~\ref{sec:3-1}. 
We further study the dependence of the global polarization of $\Lambda$ hyperons on the tilted QGP geometry and the initial velocity field in Sec.~\ref{sec:3-2}. 
The global polarization generated by different initial condition models -- the tilted Glauber model, AMPT, and SMASH -- are compared in Sec.~\ref{sec:3-3}.
The relation between global polarization and the directed flow of $\bar{\Lambda}$ hyperons is investigated in Sec.~\ref{sec:3-4}. 
In the end, we present results for the local polarization of $\Lambda$ hyperons in Sec.~\ref{sec:3-5}  

\begin{figure}[tbp!]
\begin{center}
\includegraphics[width=0.8\linewidth]{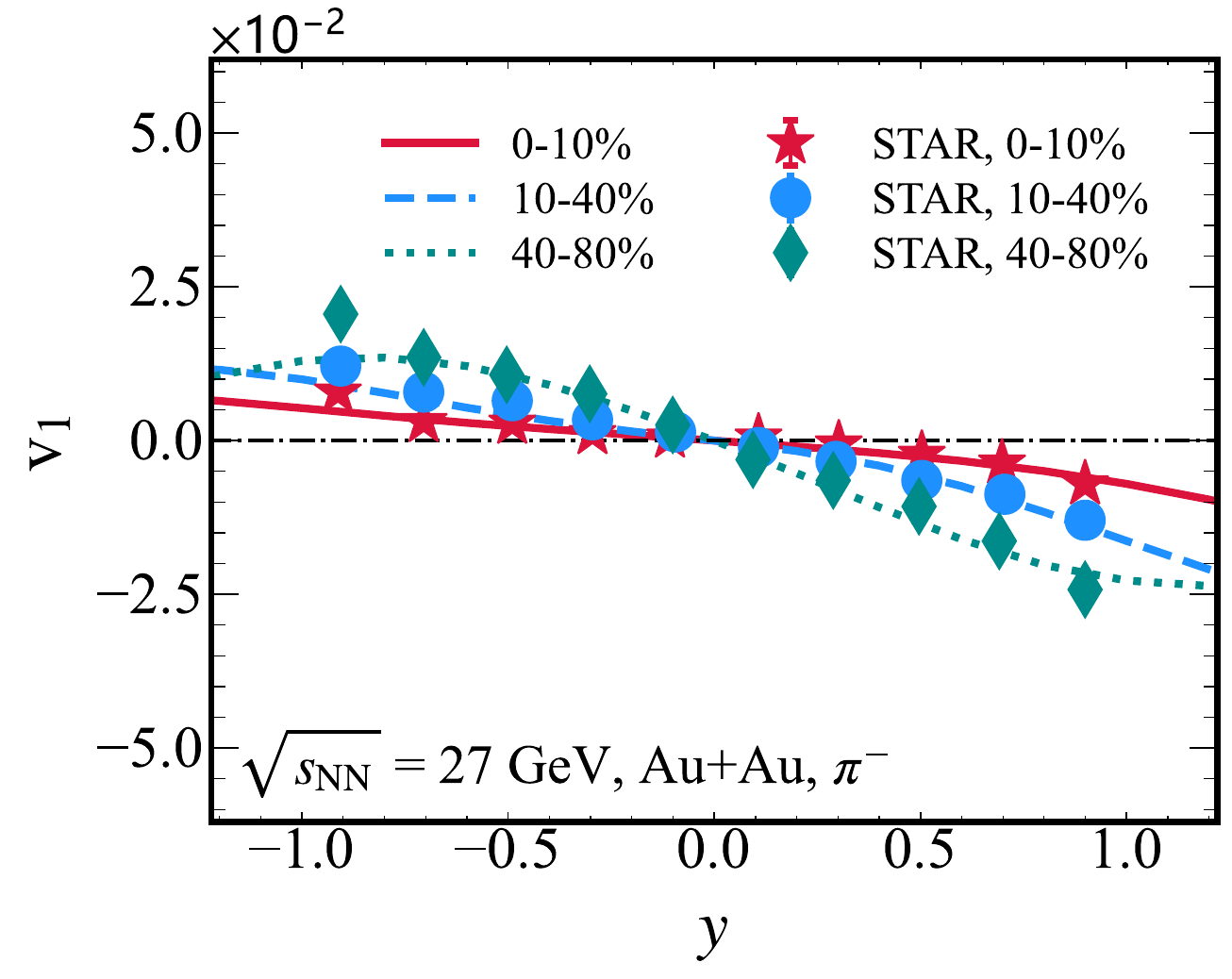}
\includegraphics[width=0.8\linewidth]{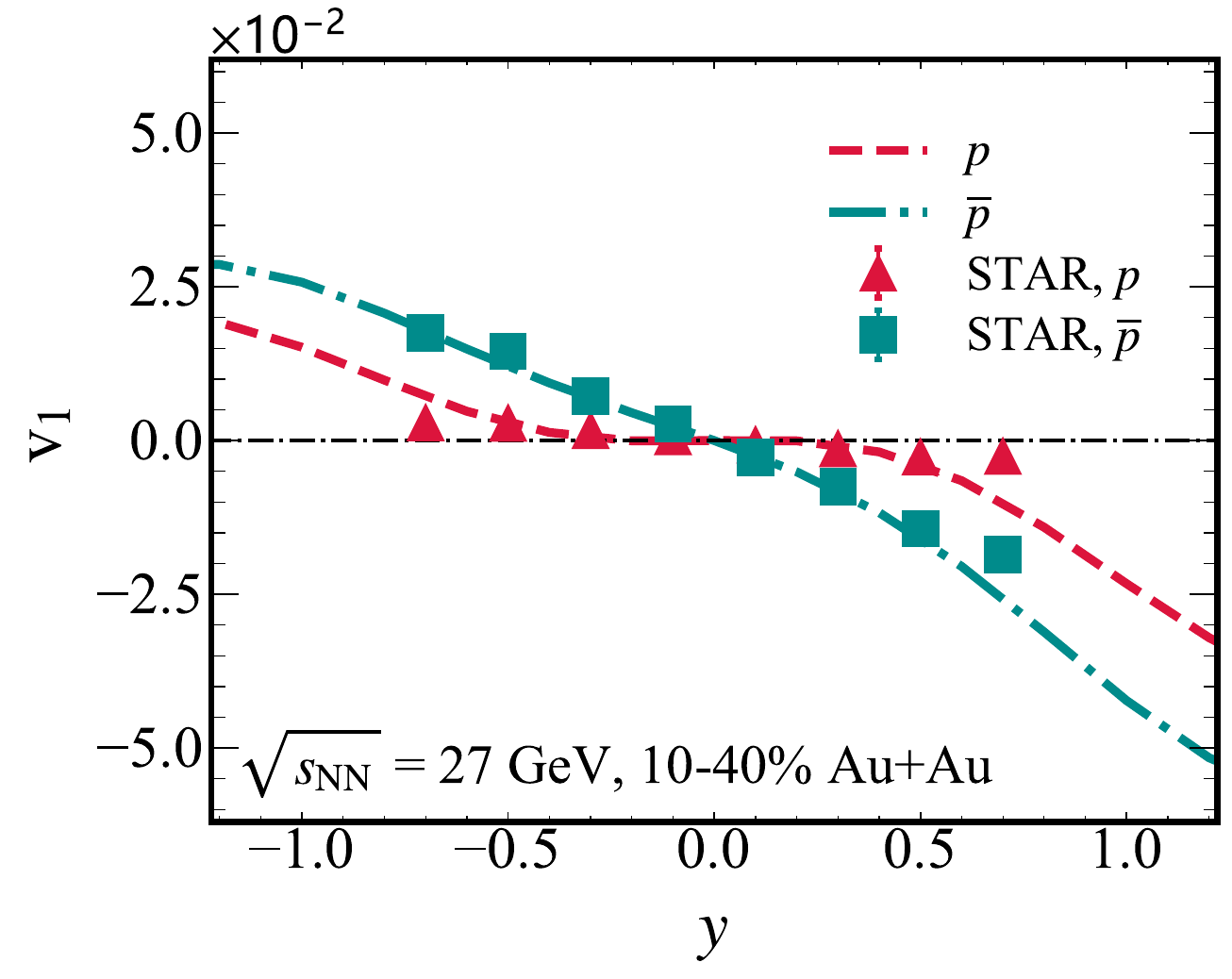}
\includegraphics[width=0.8\linewidth]{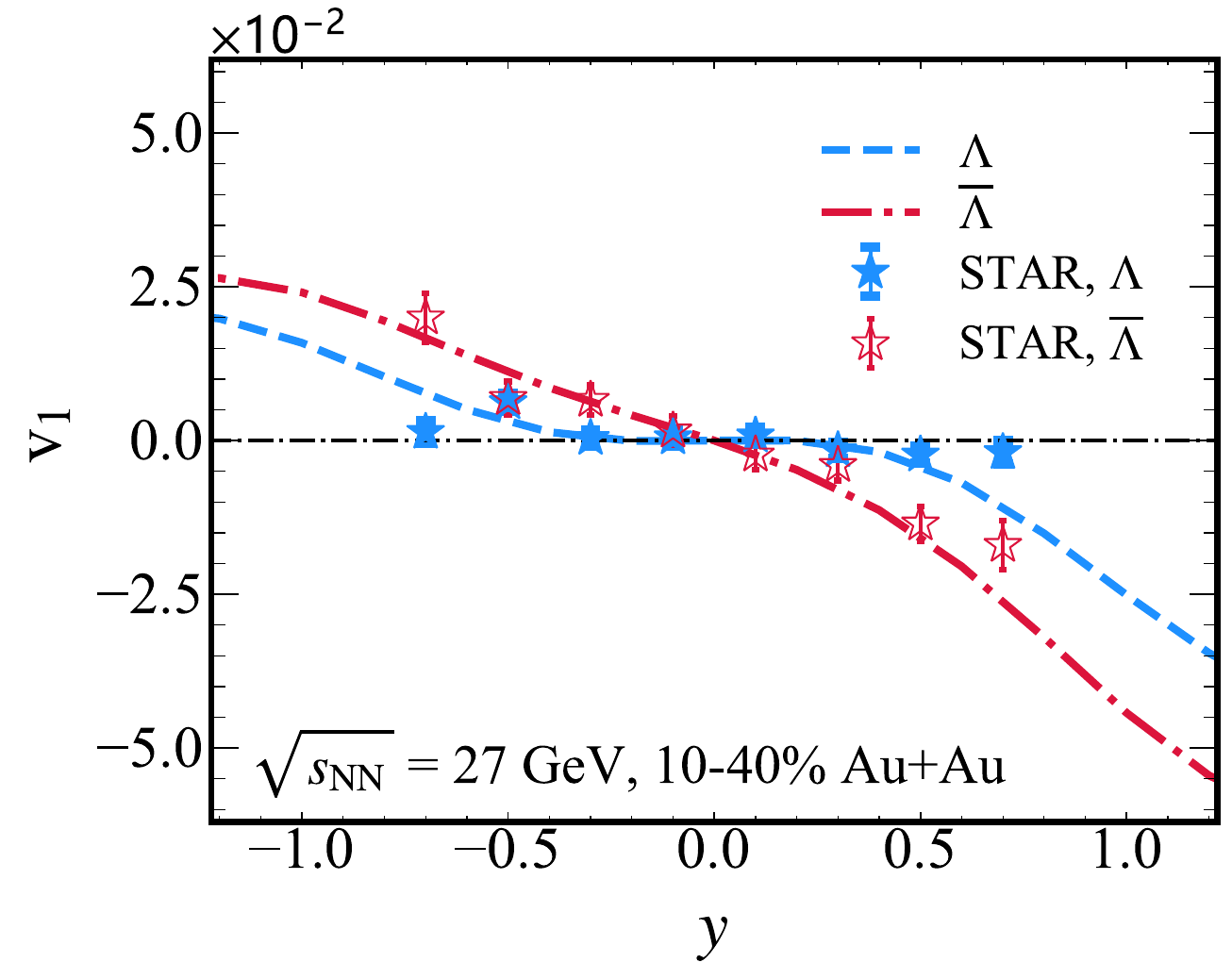}
\end{center}
\caption{(Color online) The rapidity dependence of the directed flow coefficients of $\pi^{-}$ (upper panel), $p$ and $\bar{p}$ (middle panel), and $\Lambda$ and $\bar{\Lambda}$ (lowr panel) in Au+Au collisions at $\snn=27$ GeV, compared between hydrodynamic calculation and the STAR data~\cite{STAR:2014clz}.}
\label{f:v1}
\end{figure}

%-------- Sec 3-1 ------
\subsection{Directed flow of identified particles and global polarization of $\Lambda$ hyperons}
\label{sec:3-1}

\begin{figure*}[tbh]
\includegraphics[width=0.31\textwidth]{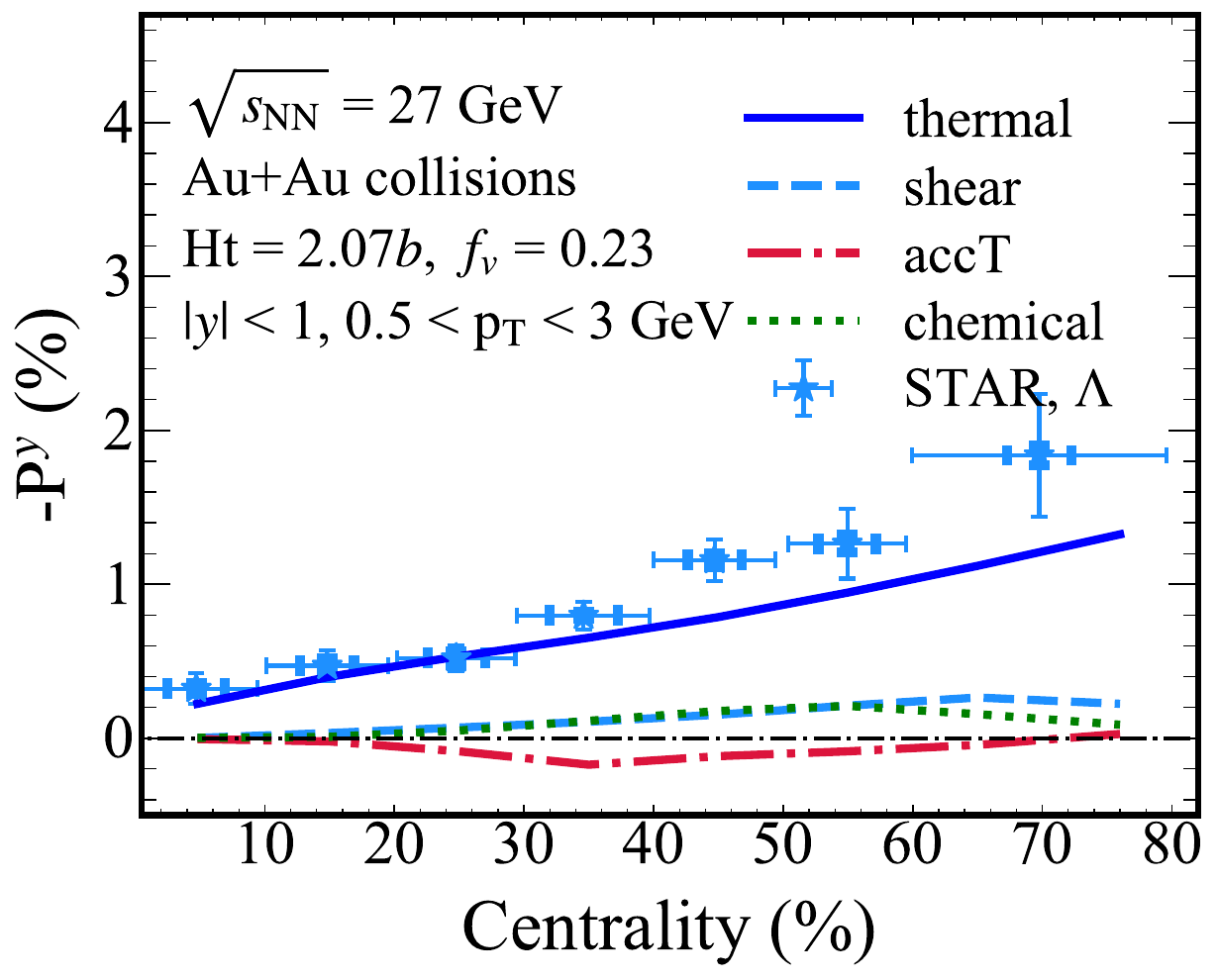}
\includegraphics[width=0.31\textwidth]{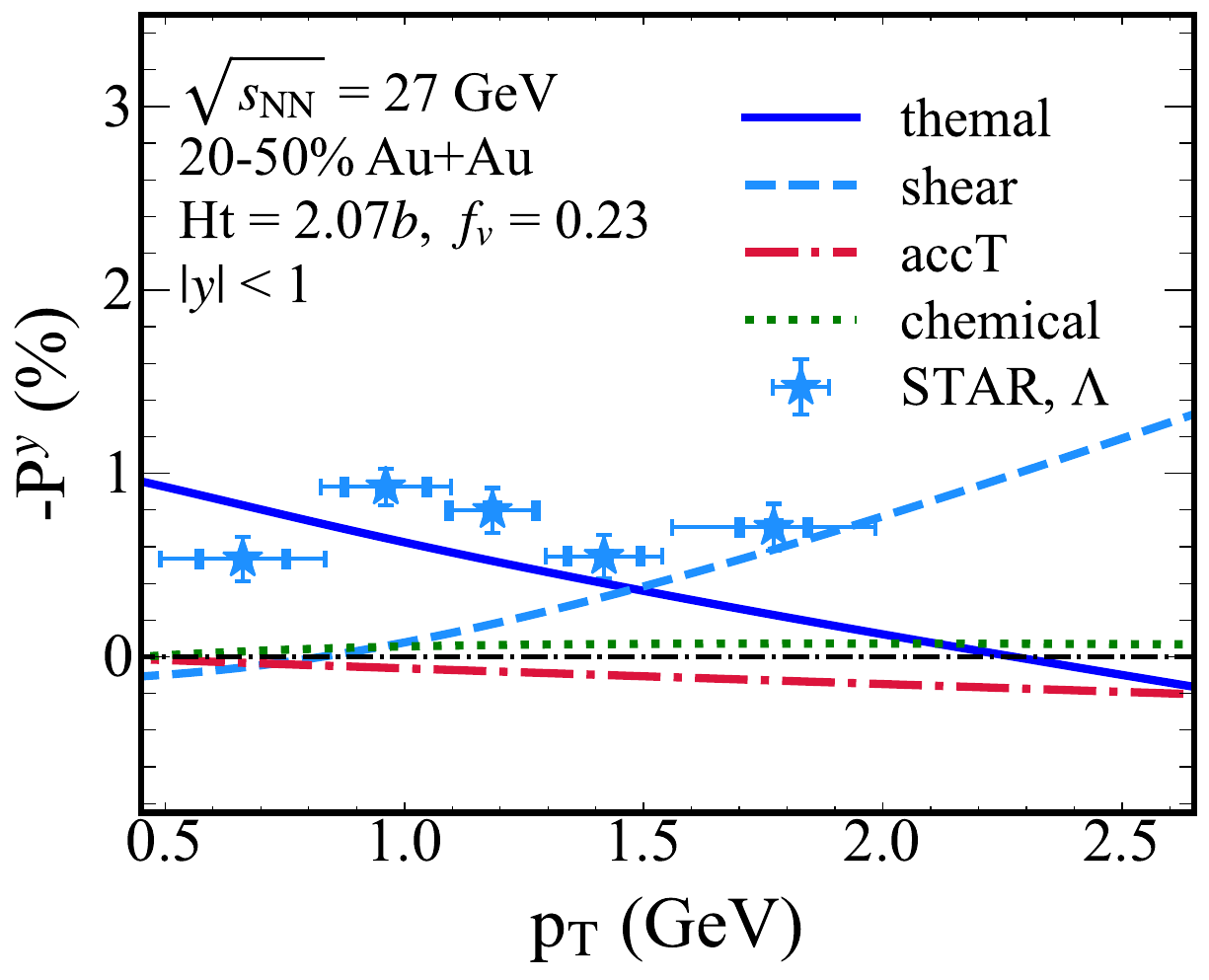}
\includegraphics[width=0.31\textwidth]{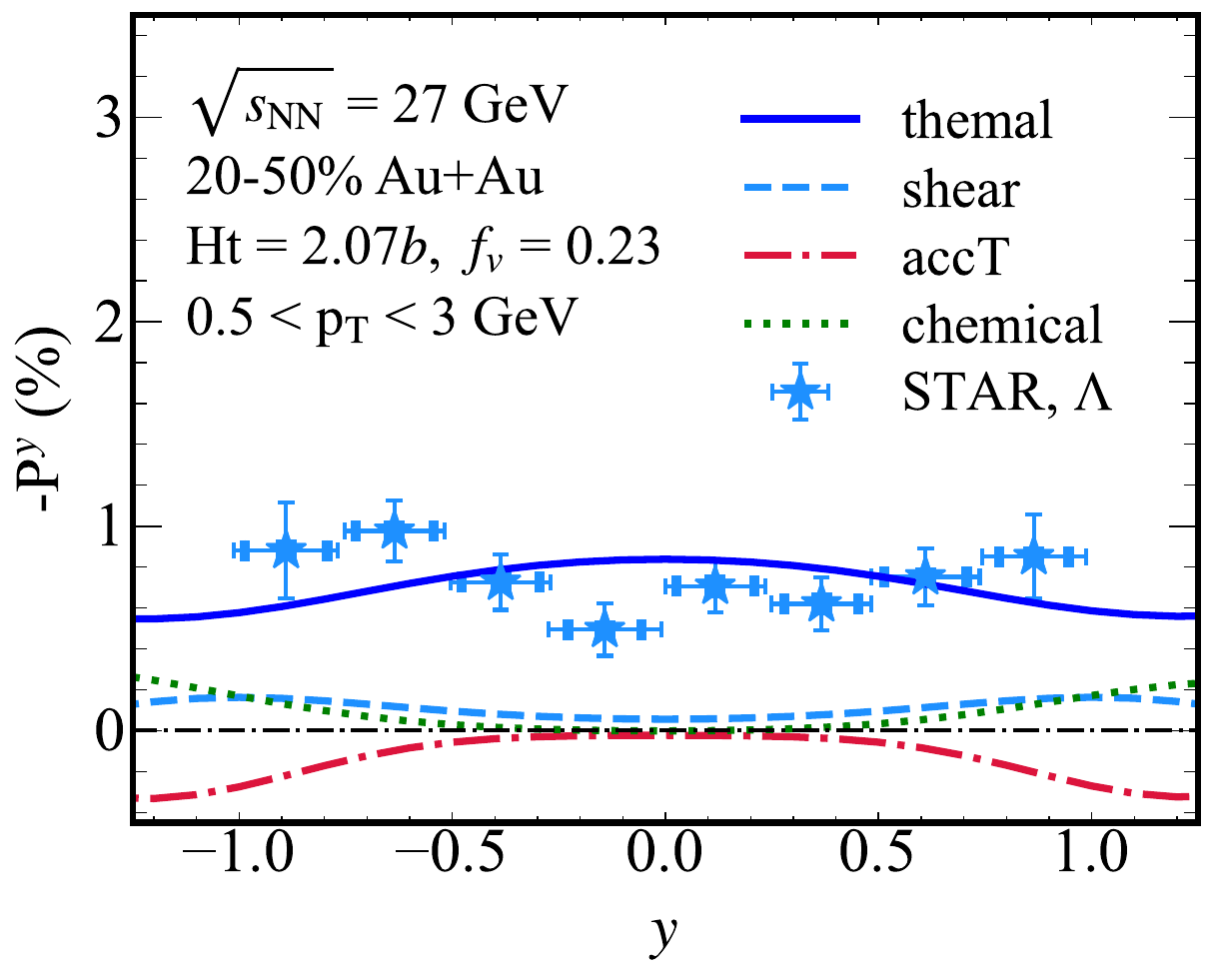}
\vspace{-4pt}
\includegraphics[width=0.31\textwidth]{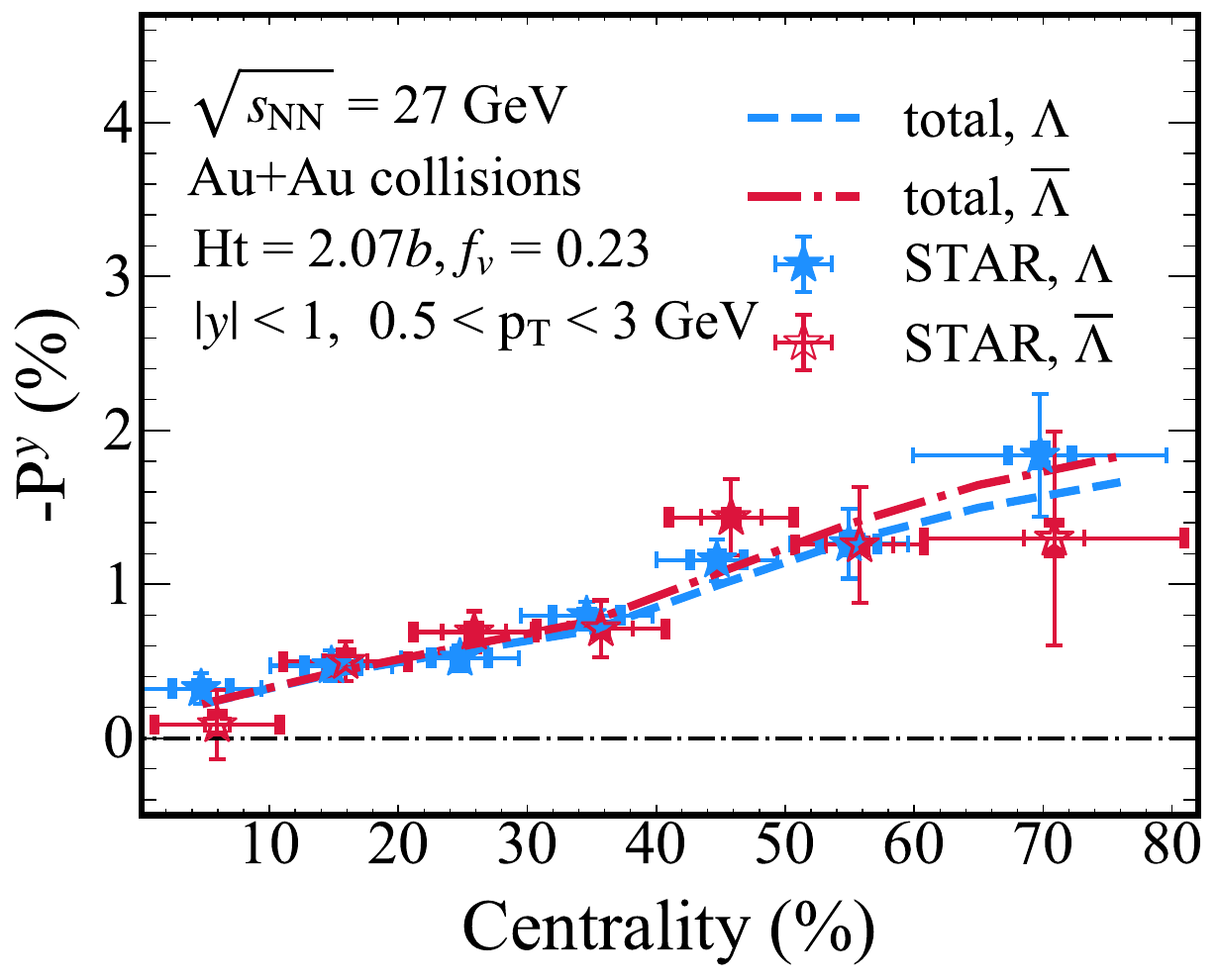}
\includegraphics[width=0.31\textwidth]{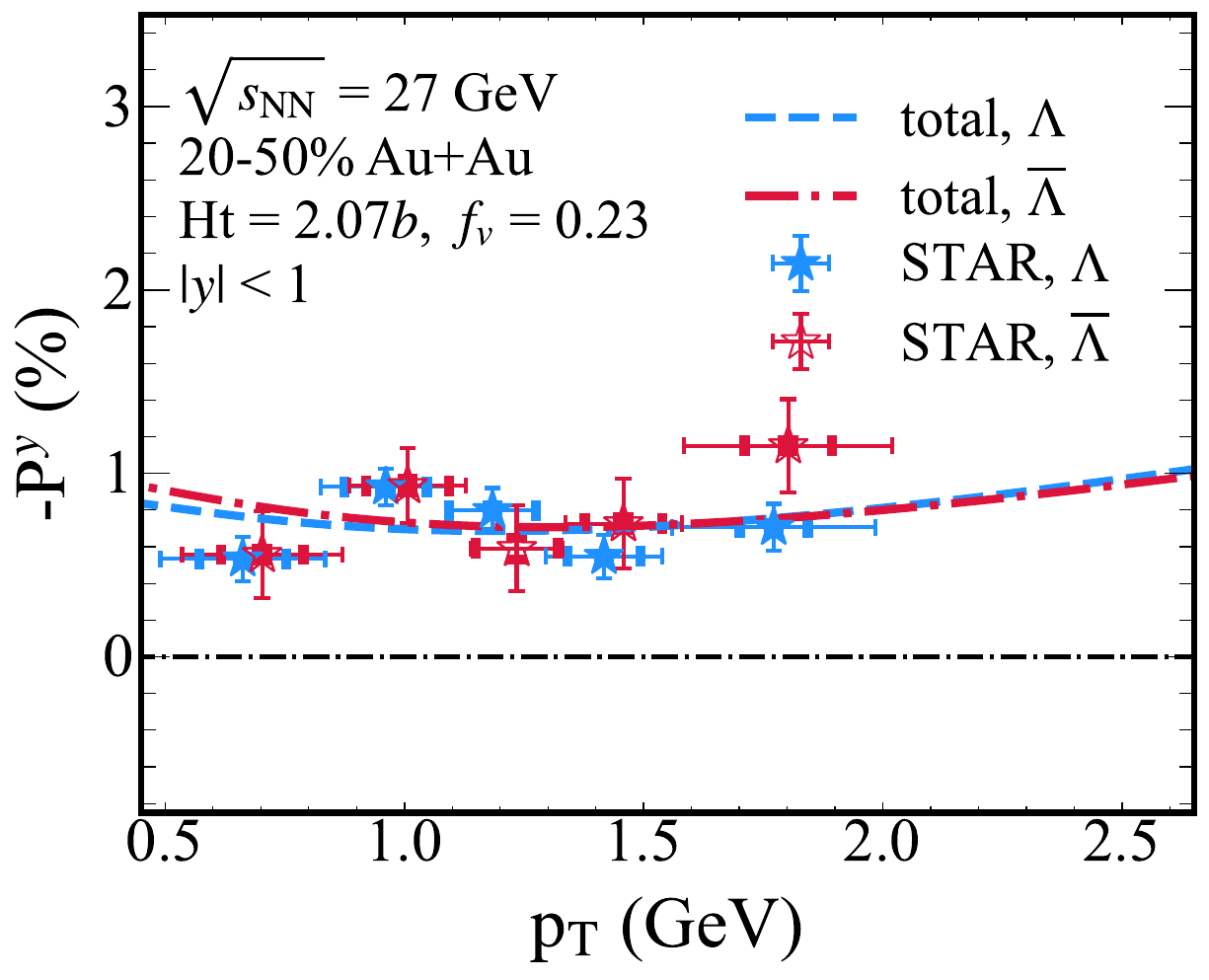}
\includegraphics[width=0.31\textwidth]{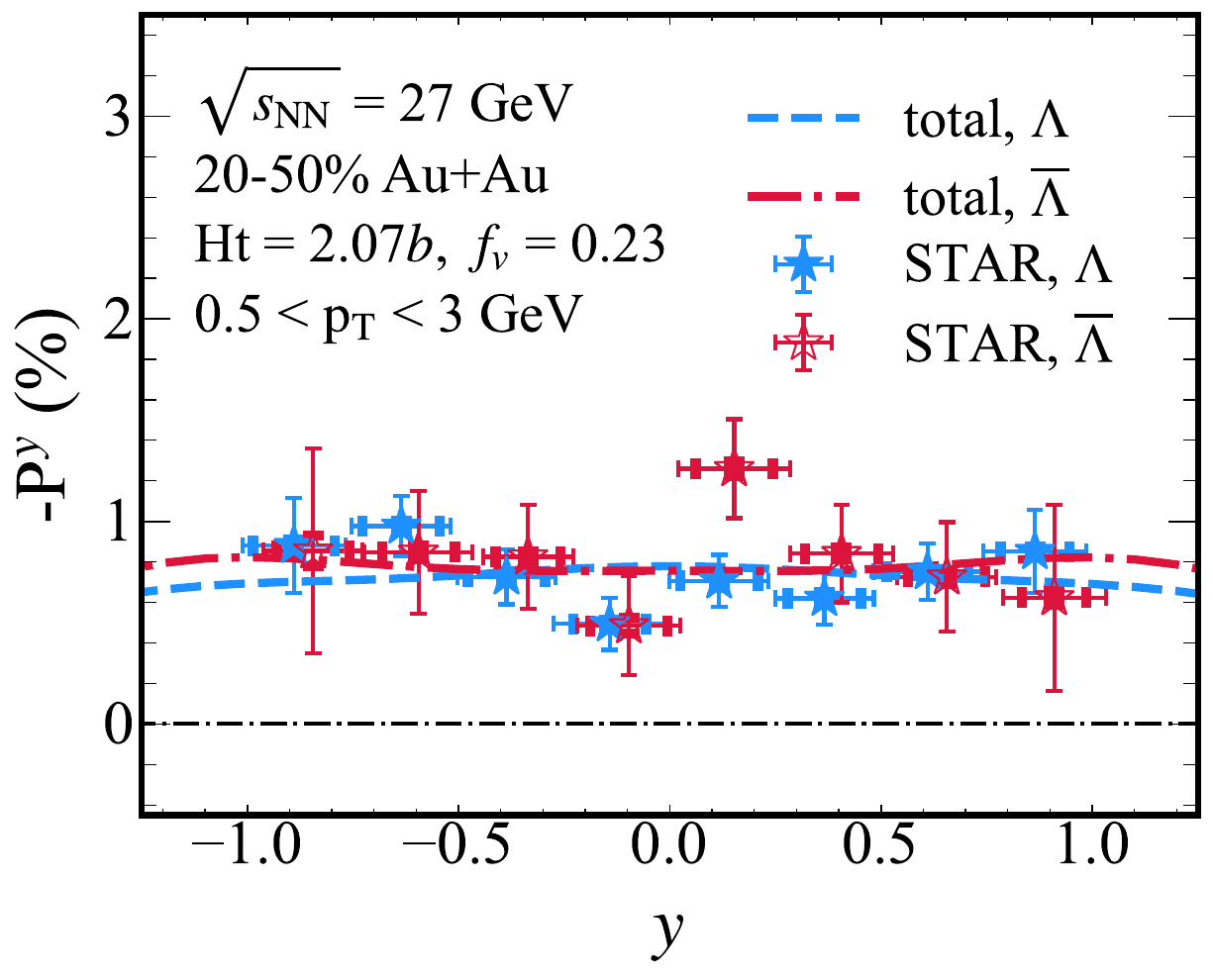}
\caption{(Color online) The global polarization of $\Lambda$ hyperons, $-P^{y}$, as functions of centrality (left column), transverse momentum (middle column) and rapidity (right column) in Au+Au collisions at $\snn=27$~GeV. The upper row compares different contributions to the $\Lambda$ polarization, while the lower panel compares between $\Lambda$ and $\bar{\Lambda}$. The experimental data are taken from the STAR Collaboration~\cite{STAR:2023nvo}.}
\label{f:global_py}
\end{figure*}

We start with validating our model setup by comparing the directed flow of identified hadrons and global polarization of $\Lambda(\bar{\Lambda})$ hyperons between our calculation and the STAR data~\citep{STAR:2017ieb} in Figs.~\ref{f:v1}-\ref{f:global_py}.

%The first-order Fourier coefficient of the azimuthal distribution of particle momentum, known as the rapidity-odd directed flow ($v_{1}$)~\cite{Voloshin:1994mz,Bilandzic:2010jr,STAR:2004jwm,STAR:2014clz,STAR:2017okv,STAR:2019clv,ALICE:2019sgg,STAR:2019vcp}, is one of the most popular and useful observables in analyzing the QGP size and properties in energetic collisions~\cite{Gyulassy:1981nq,Gustafsson:1984ka,Lisa:2000ip,PHENIX:2003qra,ALICE:2010suc,CMS:2012zex,Voloshin:1994mz,Nara:2016phs,Chatterjee:2017ahy,Singha:2016mna,Zhang:2018wlk,Guo:2017mkf,Parida:2022lmt,Bozek:2022svy,Parida:2022zse,Parida:2022ppj,Du:2022yok,Jing:2023zrh,Sun:2023adv,Parida:2023ldu,Nakamura:2022idq}. 

The directed flow coefficient $v_1$ can be extracted as the first-order Fourier coefficient of the azimuthal distribution of particle momentum as
\begin{equation}
\begin{aligned}
v_{1}(y)=\langle\cos(\phi-\Psi_{1})\rangle=\frac{\int\cos(\phi-\Psi_{1})\frac{dN}{dy d\phi}d\phi}{\int\frac{dN}{dy d\phi}d\phi},
\label{eq:v1}
\end{aligned}
\end{equation}
where $\Psi_{1}$ is the first order event plane angle of a nucleus-nucleus collision.
Due to the use of a smooth initial condition of the energy density and baryon number density, effects of event-by-event fluctuations have not been taken into account.
As a result, the event plane coincides with the spectator plane, which can be identified using deflected neutrons measured at large rapidity.
%We will investigate the impact of the initial-state fluctuations on the final-state hadron $v_1$ in our future research.

In Fig.~\ref{f:v1}, we first present the $v_1$ of different species of hadrons as a function of rapidity in Au-Au collisions at $\snn=27$~GeV. The transverse momentum range $0<p_{\rm{T}}<3.0$~GeV of these hadrons is used for the analysis. In the upper panel, we show the $v_1$ of $\pi^{-}$ in three different centrality regions. By using a linear dependence $H_\mathrm{t}=2.07 b/\textrm{fm}$ between the tilt parameter and the impact parameter in Eq.~(\ref{eq:mnccnu}), a reasonable centrality dependence of the pion $v_1$ can be obtained. Using the same model setup, we present the $v_1$ of protons and anti-protons in the middle panel for a given centrality bin. As discussed in Refs.~\cite{Bozek:2022svy,Jiang:2023fad}, introducing the tilted geometry for the net baryon density provides a satisfactory description of the splitting of $v_1$ between $p$ and $\bar{p}$. Similarly, our model results on the $v_1$ of $\Lambda$ and $\bar\Lambda$ are also consistent with the STAR observation~\cite{STAR:2014clz}, as shown in the lower panel of Fig.~\ref{f:v1}. 
  
In Fig.~\ref{f:global_py}, we present the global polarization of hyperons along the out-of-plane direction, $-P^{y}$, analyzed within the kinematic region of $p_\text{T} \in [0.5~\text{GeV},~3.0~\text{GeV}]$ and $y\in [-1,~1]$. In the upper panels, we compare different contributions, i.e., different terms in Eq.~(\ref{eq:totS}), to the polarization of $\Lambda$ as functions of (from left to right) centrality, transverse momentum and rapidity, respectively. One observes that after integrating over $p_\mathrm{T}$, the thermal vorticity is the dominant contributor to the global polarization of $\Lambda$ across different centralities and rapidities (left and right). However, in the middle panel, it is interesting to note that opposite tends with respect to $p_\mathrm{T}$ can be seen between the thermal vorticity and shear tensor contributions: the former decreases while the latter increases as $p_\mathrm{T}$ becomes larger. Contribution from the shear term becomes non-negligible above $p_\mathrm{T}\approx 1$~GeV and even becomes dominant above $p_\mathrm{T}\approx 1.5$~GeV. Later, we will show that the $p_\mathrm{T}$ dependences of these two terms rely on the medium geometry and the longitudinal flow field of the QGP.

In the lower panels of Fig.~\ref{f:global_py}, we combine contributions from the four terms (thermal, shear, accT, and chemical) and present the global polarization ($-P^{y}$) of both $\Lambda$ and $\bar{\Lambda}$ as functions of centrality, transverse momentum and rapidity. Our model calculation provides a satisfactory description of the hyperon polarization compared to the STAR data~\cite{STAR:2023nvo}. Only a minor difference is observed between $\Lambda$ and $\bar{\Lambda}$, which results from the chemical term contribution to $-P^{y}$. In addition, due to the opposite $p_\mathrm{T}$ dependences between thermal and shear contributions (middle panel in the upper row), their combination leads to a non-monotonic dependence of $-P^{y}$ on $p_\mathrm{T}$ (middle panel in the lower row). This feature can be examined with more precise data in the future, and provide more stringent constraints on different components of hyperon polarization. With these validations of our model calculation, we will explore the dependence of hyperon polarization on the medium profiles and its relation with the directed flow in the rest of this work.

%One finds that the integration of different components from our model creates a satisfactory account of the $\Lambda(\bar{\Lambda})$ hyperons observed in 20-50\% centrality at $\snn$= 27 GeV, spanning a range of centrality classes, transverse momentum, and rapidity. The $-P^{y}$ against transverse momentum show a non-monotonic pattern attributable to the tilted QGP geometry. 

%We note here that due to the smaller splitting of $-P^{y}$ observed between $\Lambda$ and $\bar{\Lambda}$ hyperons, as same as illustrated in~\cite{STAR:2023ntw}, our subsequent study will exclusively focus on investigating the polarization of $\Lambda$ or $\Lambda$ hyperons. This approach ensures consistency and coherence throughout our research.

% -------- Sec 3-2 ------

\subsection{Effects of the initial QGP geometry and longitudinal flow on global polarization}
\label{sec:3-2}

\begin{figure}[tbp!]
\begin{center}
\includegraphics[width=0.45\textwidth]{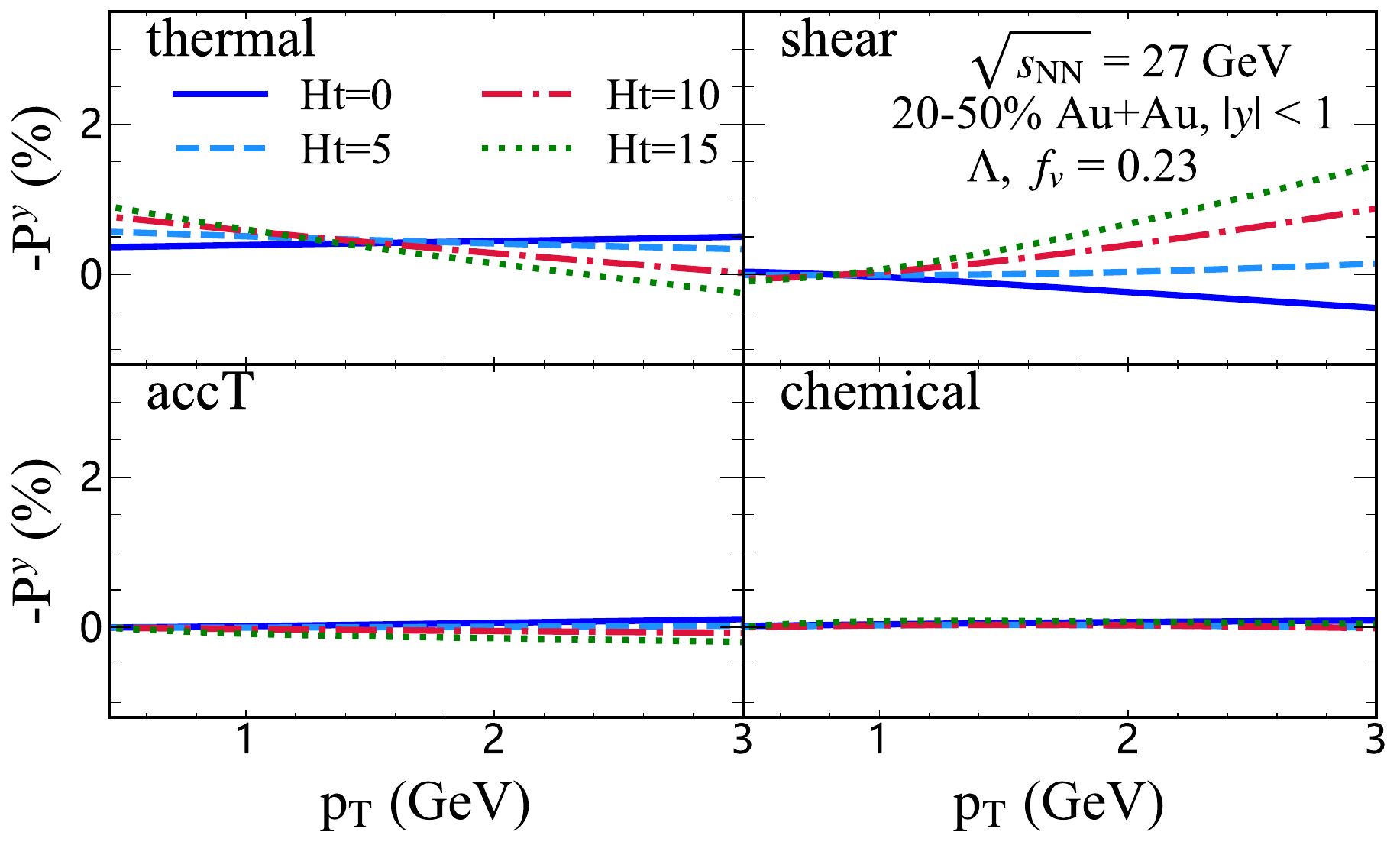}
\includegraphics[width=0.45\textwidth]{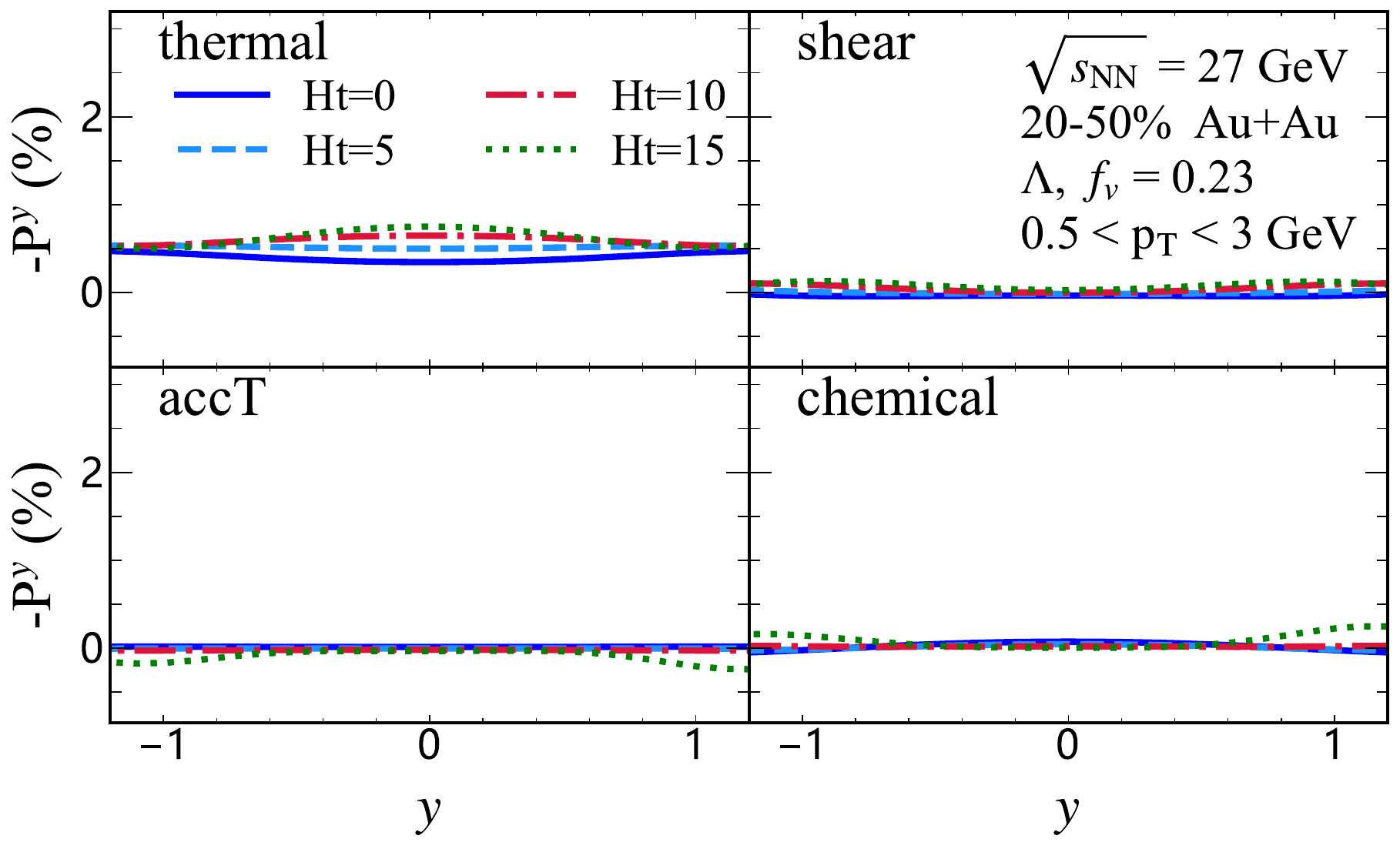}
\end{center}
\caption{(Color online) Effects of the tilted geometry of the QGP (the $H_\text{t}$ parameter) on the transverse momentum (upper plot) and rapidity (lower plot) dependences of the $\Lambda$ polarization, compared between different terms contributing to the global polarization. 
The initial longitudinal velocity field parameter is fixed at $f_{v}= 0.23$.
%Results from our model calculation are compare to the STAR data~\cite{STAR:2023ntw}.
}
\label{f:pol_ht}
\end{figure}

\begin{figure}[tbp!]
\begin{center}
\includegraphics[width=0.45\textwidth]{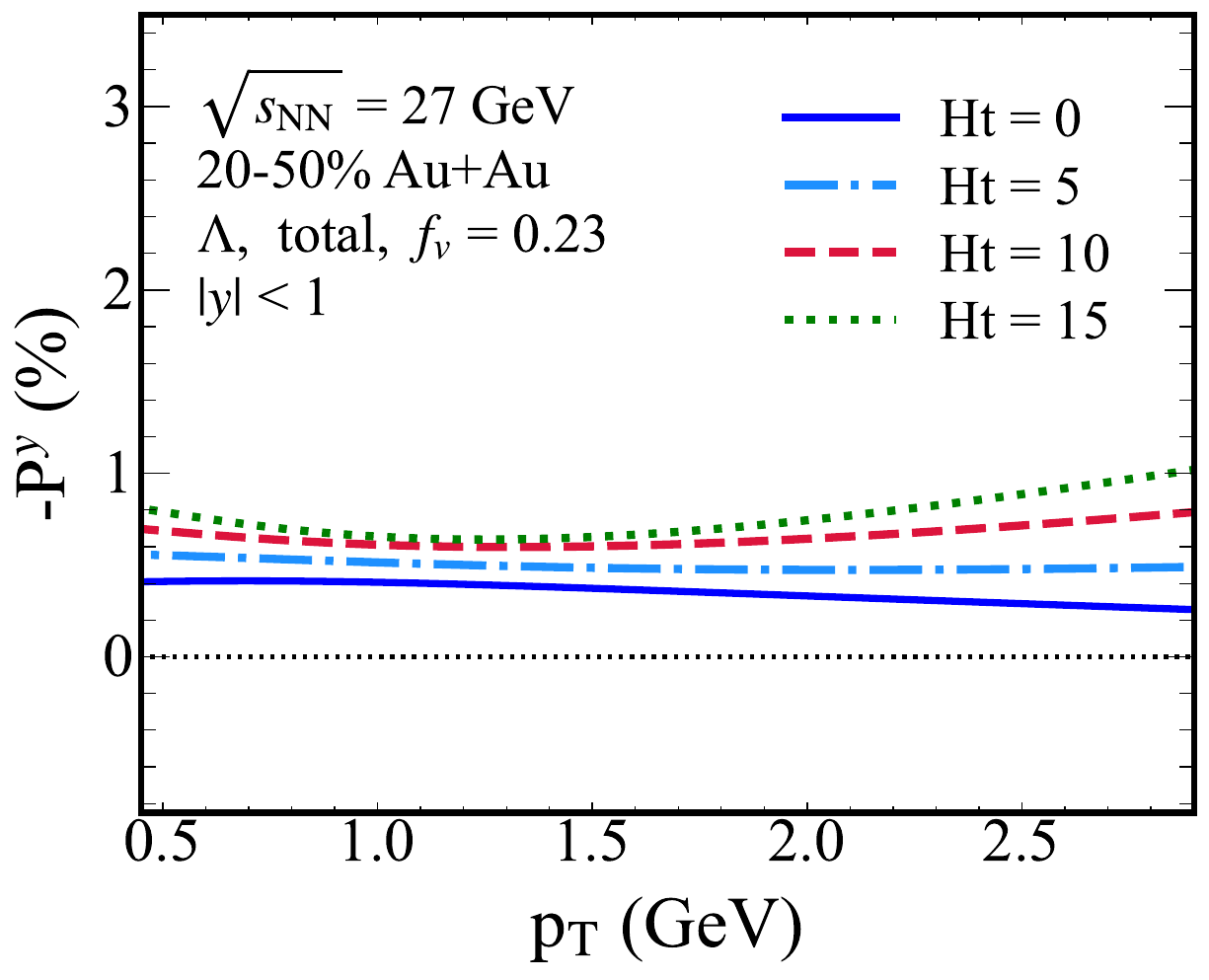}
\includegraphics[width=0.45\textwidth]{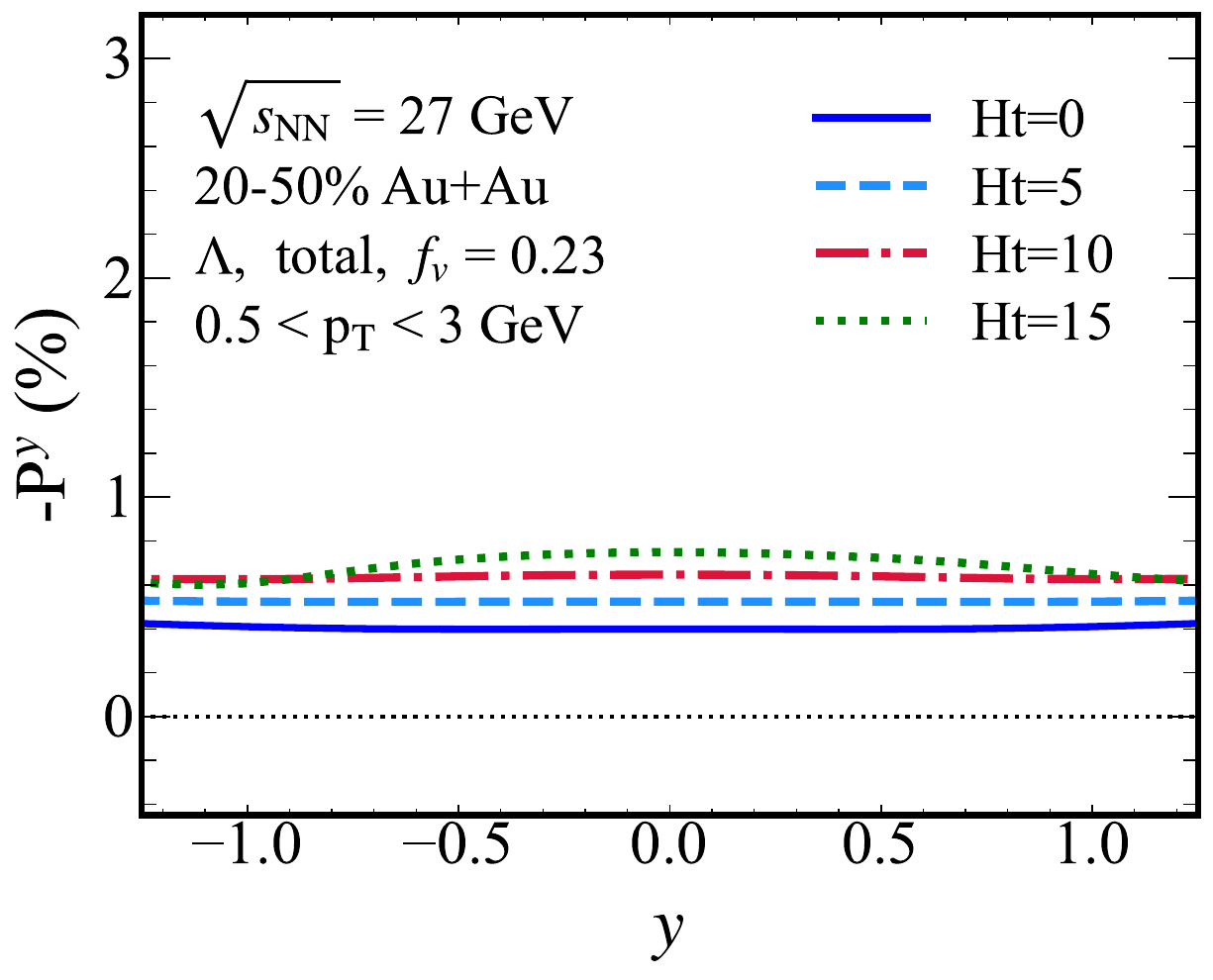}
\end{center}
\caption{(Color online) Effects of the tilted geometry of the QGP (the $H_\text{t}$ parameter) on the transverse momentum (upper panel) and rapidity (lower panel) dependences of the $\Lambda$ polarization. The initial longitudinal velocity field parameter is fixed at $f_{v}= 0.23$.}
\label{f:pol_ht_total}
\end{figure}

In this subsection, we implement a detailed analysis on how the initial geometry and longitudinal flow profiles of the QGP affect the global polarization of $\Lambda$ hyperons.

In Fig.~\ref{f:pol_ht}, we first fix the initial longitudinal flow velocity field with $f_v=0.23$ and study how the tilt of the QGP geometry influences different components of $\Lambda$ polarization. The upper plot shows the global polarization as a function of $p_\mathrm{T}$. And in each panel, we study how the $H_{\textrm{t}}$ parameter affects each contribution -- thermal, shear, accT, and chemical -- to the $\Lambda$ polarization. As $H_{\textrm{t}}$ increases from 0 to 15, one observes the slope of $-P^{y}(p_\mathrm{T})$ decreases from positive to negative values in the thermal vorticity term, while increases from negative to positive values in the shear tensor term. This could be understood with the $-u_\beta\partial_\alpha T/T^2$ component in the $S_\text{thermal}^\mu$ term and the $u_\beta/T$ component in the $S_\text{shear}^\mu$ term, which are both amplified with a more asymmetric medium and lower temperature at mid-rapidity when $H_\text{t}$ increases.
%{\color{red} This is expected because a more tilted fireball induced a more asymmetry and lower temperature in the transverse plane at the middle rapidity.} 
Consequently, the non-monotonic dependence of their combination on $p_\mathrm{T}$ may provide additional constraint on the medium geometry if the experimental data becomes sufficiently precise. 
%{\color{blue} (Can we add a few sentences to explain why increasing $H_\text{t}$ gives opposite change of these two terms?)} 
Little impact from $H_\text{t}$ has been found on the $\Lambda$ polarization from the fluid acceleration (accT) term and the SHE (chemical) term. A similar investigation is conducted in the lower plot of Fig.~\ref{f:pol_ht}, where the $\Lambda$ polarization is studied as a function of rapidity. As the value of $H_\text{t}$ increases from 0 to 15, the dip structure of the $\Lambda$ polarization at mid-rapidity from the thermal vorticity term gradually transits into a peak structure. The value of this global polarization near $y=0$ is enhanced from 0.40 to 0.73. For the other three terms of global polarization, impact of this tilted deformation of the QGP appears small.

%In the lower panel of Fig.~\ref{f:pol_ht}, we present the rapidity ($y$) dependence of the global polarization of $\Lambda$ hyperons along the out-of-plane direction, resulting from various components with $H{_\textrm{t}}$ values of 0, 5, 10, and 15. Our analysis reveals that an increase in $H{_\textrm{t}}$ leads to a substantial enhancement in the thermal vorticity-induced polarization (thermal) from 0.40 to 0.73 at $|y|<0.5$. In addition, we observe the magnitudes of the shear-induced polarization (shear), fluid acceleration-induced polarization (accT) and the SHE-induced polarization (chemical) change indistinguishably.

In Fig.~\ref{f:pol_ht_total}, we combine contributions from the four terms above and present the total value of $\Lambda$ polarization as functions of both $p_\mathrm{T}$ (upper panel) and $y$ (lower panel). When the $f_v$ parameter is fixed at 0.23, one observes an enhancement in the value of $-P^y$ as one increases the tilt parameter $H_\text{t}$. Meanwhile, a clear non-monotonic behavior of polarization with respect to $p_\text{T}$ appears when $H_\text{t}$ is sufficiently large, which may serve as a signature of the tilted geometry of the QGP fireball. 

%It is evident that the magnitude of $P^{y}$ is amplified with increasing $H\rm{_t}$ from 0.0 to 15. However, this amplification diminishes due to the contribution from the shear induced polarization when 0 $<$ $p_{\rm{T}}$ $<$ 2.5 GeV. The observed non-monotonic behavior of the transverse momentum distribution of the global polarization $-P^{y}$ may serve as a promising indicator for identifying the geometry of a tilted quark-gluon plasma fireball.   
%Moreover, the lower panel of Fig.~\ref{f:pol_ht_total} presents the integrated polarization of $\Lambda$ hyperons as a function of rapidity ($y$). 
%Notably, within the rapidity region of $|y|<0.5$, the absolute value of $P^{y}$ rises with the augmentation of $H_{\textrm{t}}$.

\begin{figure}[tbp!]
\begin{center}
\includegraphics[width=0.45\textwidth]{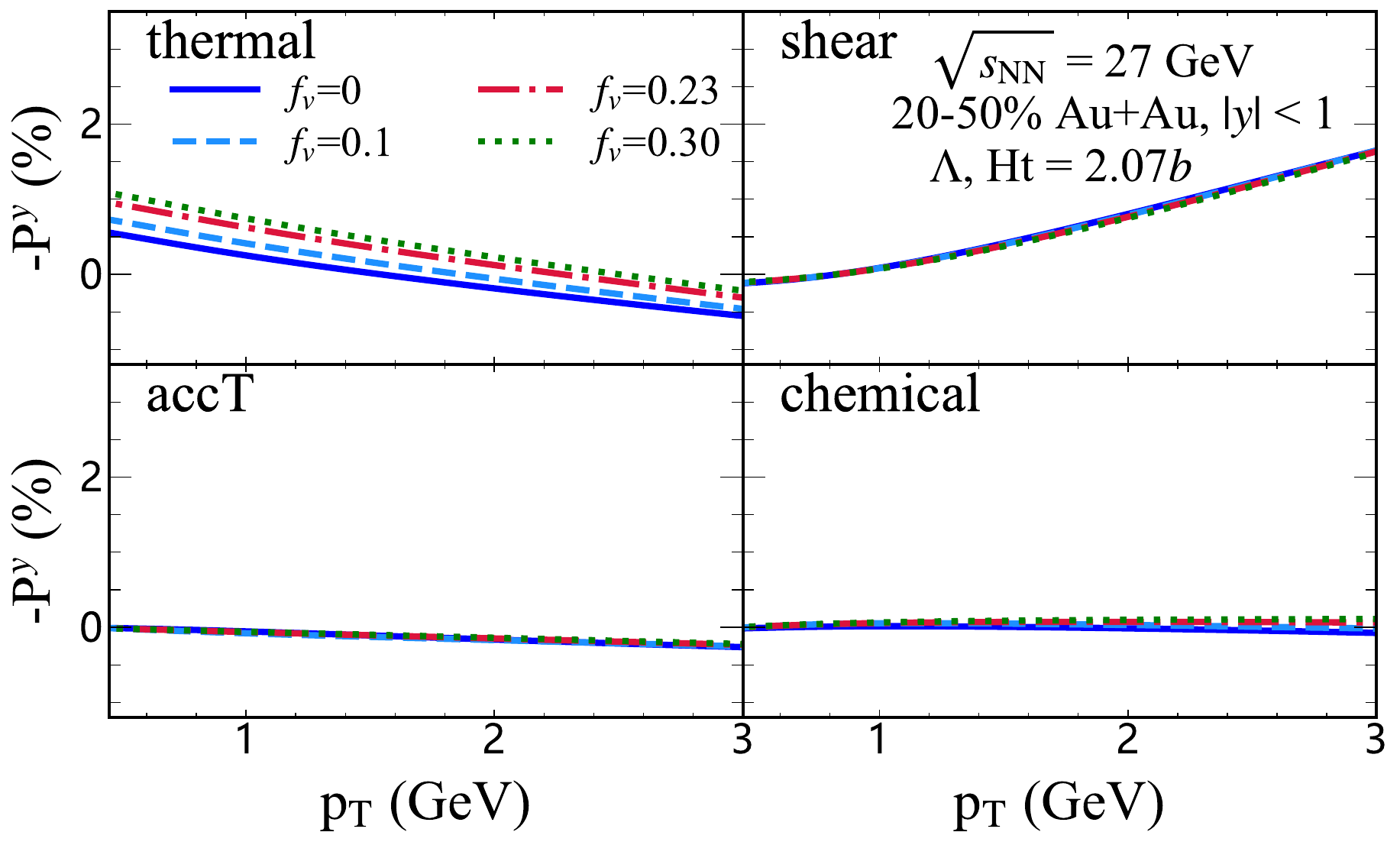}
\includegraphics[width=0.45\textwidth]{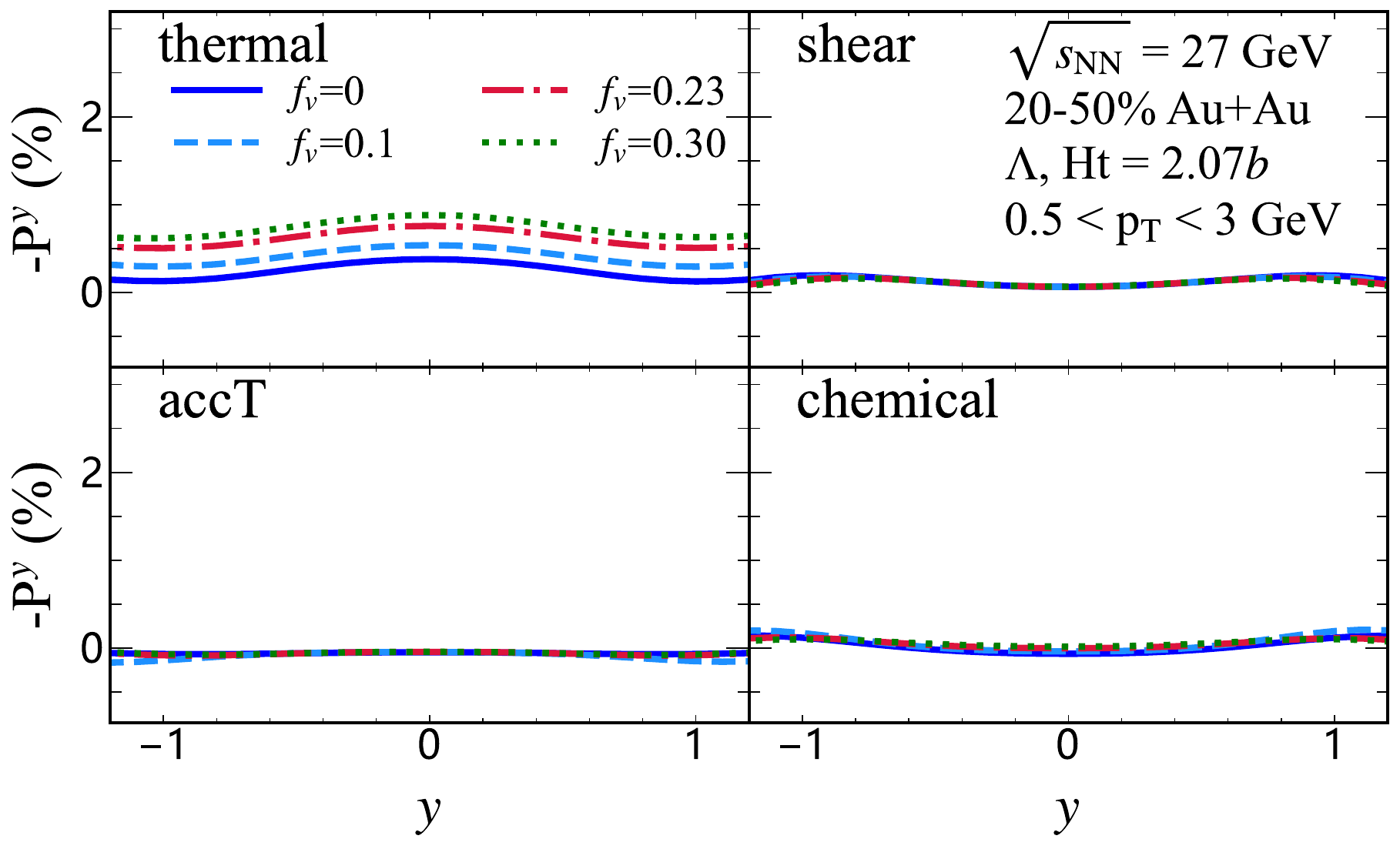}
\end{center}
\caption{(Color online) Effects of the initial longitudinal flow velocity field (the $f_v$ parameter) on the transverse momentum (upper plot) and rapidity (lower plot) dependences of the $\Lambda$ polarization, compared between different terms contributing to the global polarization. The titled geometry of the medium is fixed via $H\rm{_t} = 2.07 b/\mathrm{fm}$.}
\label{f:pol_fy}
\end{figure}

\begin{figure}[tbp!]
\begin{center}
\includegraphics[width=0.45\textwidth]{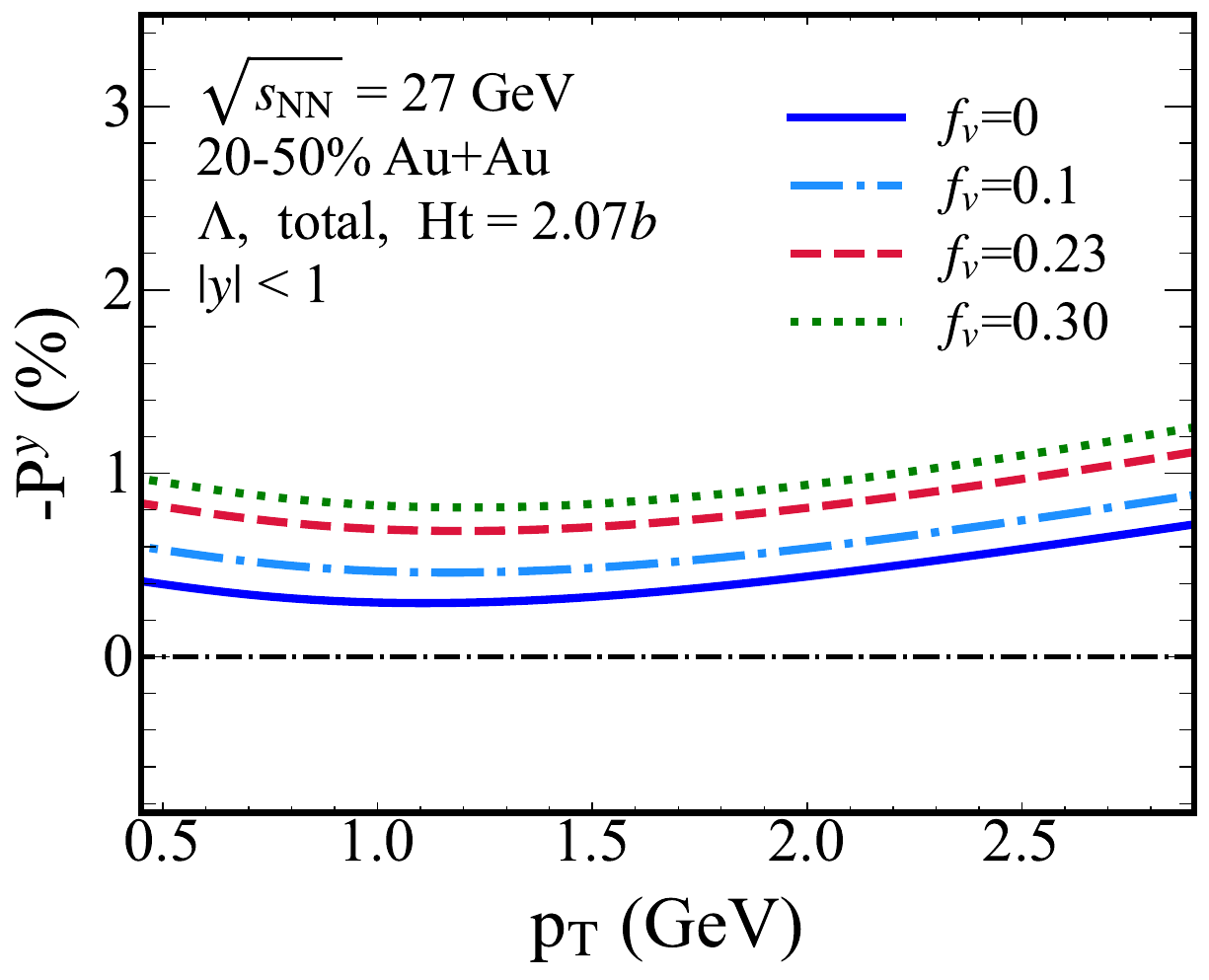}
\includegraphics[width=0.45\textwidth]{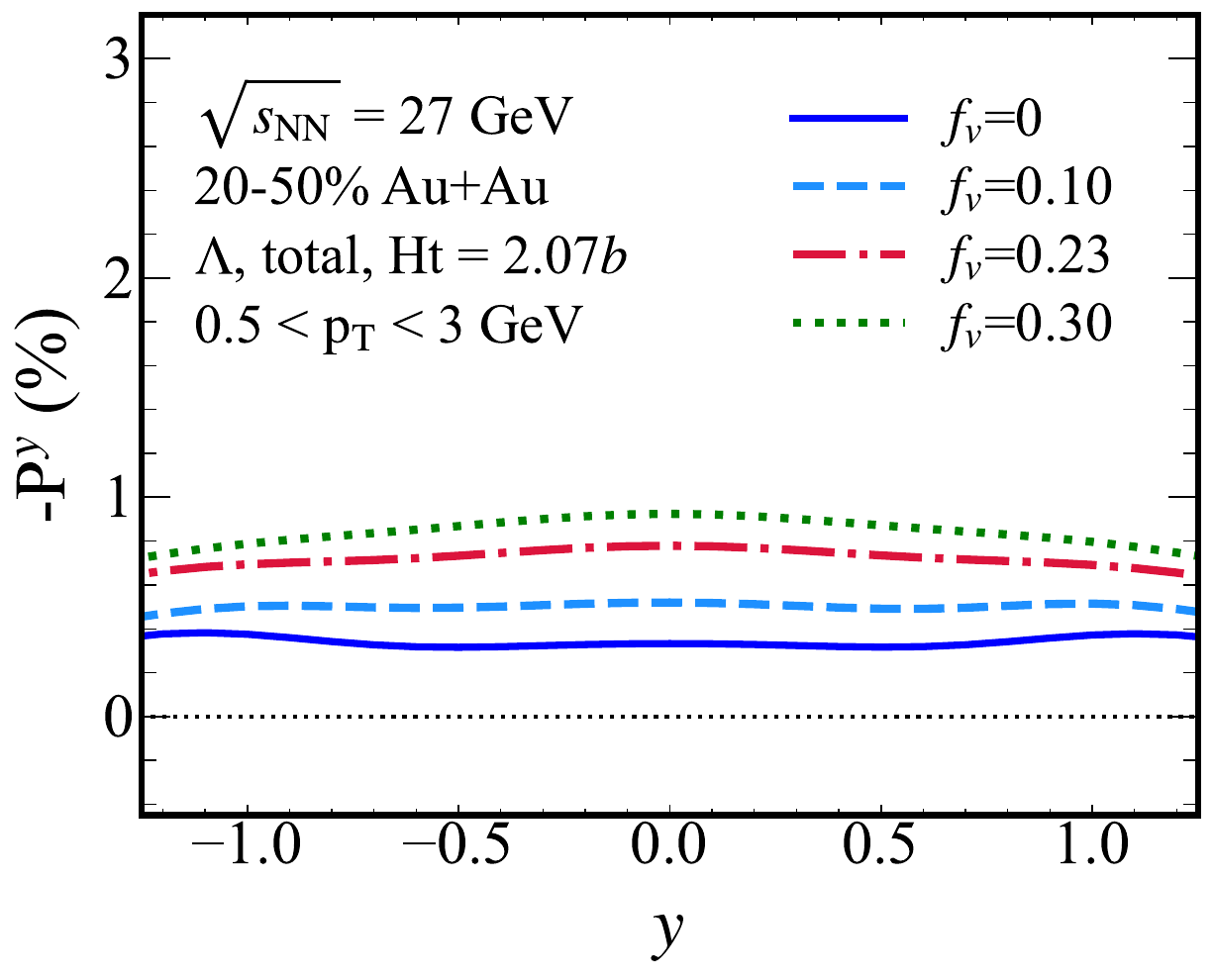}
\end{center}
\caption{(Color online) Effects of the initial longitudinal flow velocity field (the $f_v$ parameter) on the transverse momentum (upper plot) and rapidity (lower plot) dependences of the $\Lambda$ polarization. The titled geometry of the medium is fixed via $H\rm{_t} = 2.07 b/\mathrm{fm}$.}
\label{f:pol_fy_total}
\end{figure}

Similarly, we study the relation between the longitudinal flow velocity field (or $f_{v}$) and the global polarization in Figs.~\ref{f:pol_fy} and~\ref{f:pol_fy_total}. Here, we fix $H_\text{t}=2.07b/\mathrm{fm}$ for the medium geometry, which is fitted from the centrality dependence of the hadron $v_1$ earlier. In Fig.~\ref{f:pol_fy}, we present $p_\mathrm{T}$ (upper plot) and $y$ (lower plot) dependences of $-P^{y}$ for four different contributions separately. As one increases the value of $f_v$ from 0 to 0.3, an enhanced global polarization is seen from the thermal vorticity term. This can be understood with the stronger longitudinal velocity gradient deposited into the QGP when $f_v$ becomes larger, which directly increases the global vorticity of the medium and therefore the $\Lambda$ polarization. On the other hand, little variation is observed in the other three terms when we change the $f_v$ parameter. The total value of polarization is presented in Fig.~\ref{f:pol_fy_total} after contributions from the four terms are combined. When the medium geometry is fixed via $H_\text{t}=2.07b/\mathrm{fm}$, a non-monotonic $p_\mathrm{T}$ dependence of $\Lambda$ polarization can be observed in the upper panel for different values of $f_v$ applied here. Increasing the $f_v$ value significantly enhances the magnitude of polarization. As shown in the lower panel, this enhancement appears more prominent at mid-rapidity than at large rapidity. 

\subsection{Comparison between different initialization models}
\label{sec:3-3}

\begin{figure}[tbp!]
\begin{center}
\includegraphics[width=0.45\textwidth]{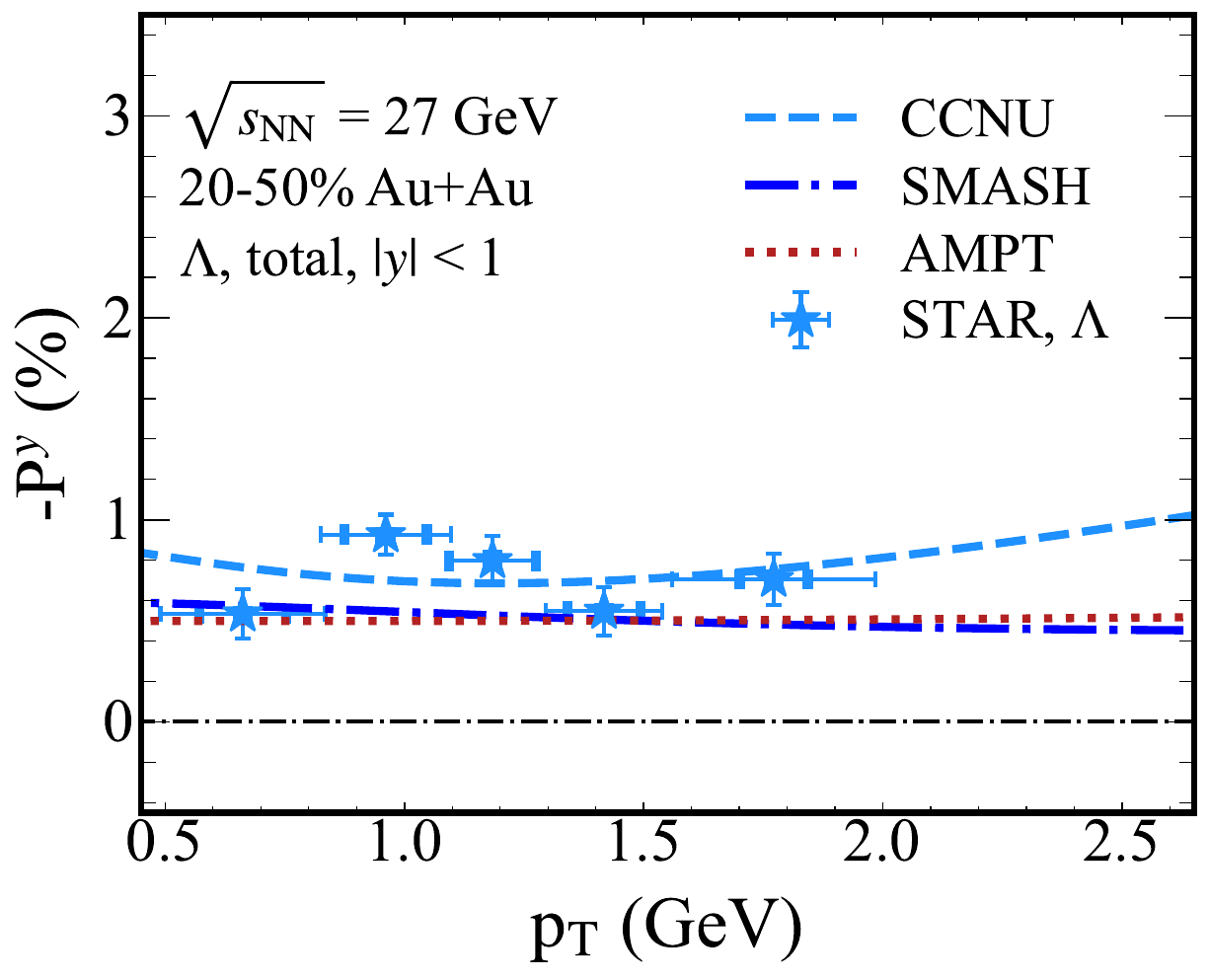}
\includegraphics[width=0.45\textwidth]{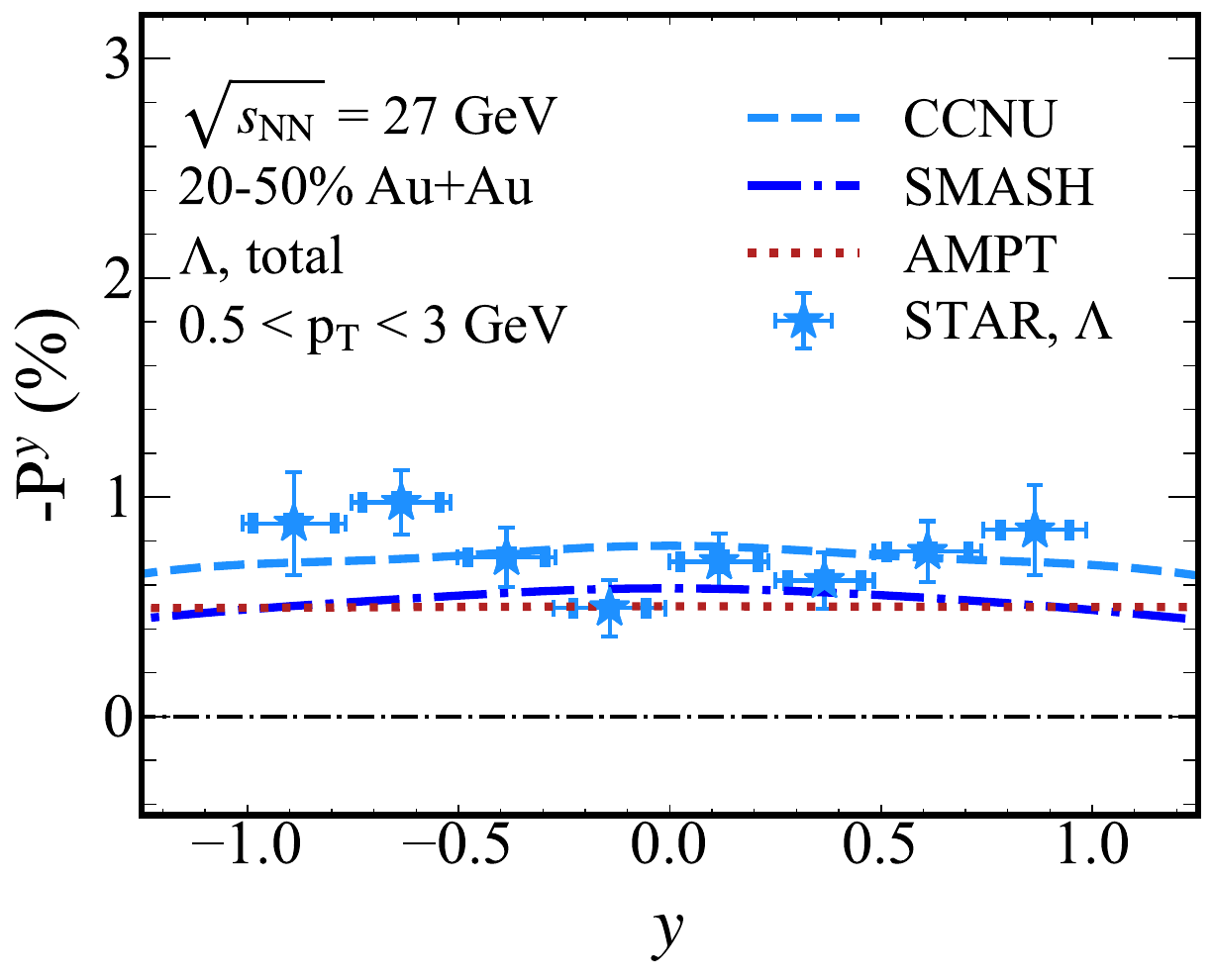}
\end{center}
\caption{(Color online) The global polarization of $\Lambda$ hyperons as functions of $p_{\text{T}}$ (upper panel) and $y$ (lower panel) in 20-50\% Au+Au collisions at $\snn=27$~GeV, compared between CLVisc hydrodynamic calculation using three different initialization methods and the STAR data~\cite{STAR:2023nvo}.
}
\label{f:ampt_smash}
\end{figure}

Constraining the initial condition from the final state hadron observables is an ongoing effort of heavy-ion programs. It has been suggested in Ref.~\cite{Wu:2022mkr} that the $\Lambda$ polarization can be affected by implementing different initialization models. Therefore, it is of great interest to investigate whether the initial condition we develop in this work introduces further impacts on polarization. In this subsection, we compare the $\Lambda$ polarization between three different initialization methods: the titled optical Glauber model described in Sec.~\ref{v1subsect2}, SMASH~\cite{Weil:2016zrk} and AMPT~\cite{Lin:2004en}. The parameters and settings of SMASH and AMPT are identical to those used in Ref.~\cite{Wu:2022mkr}. And after the CLVisc hydrodynamic evolution, these three initial conditions are able to produce comparable $p_\mathrm{T}$ spectra of charged particles. 

Shown in Fig.~\ref{f:ampt_smash} is the global polarization of $\Lambda$ in 20-50\% Au+Au collisions at $\snn=27$~GeV as functions of $p_{\rm{T}}$ (upper panel) and $y$ (lower panel), compared between CLVisc hydrodynamic calculations with three different initialization models. One can observe a larger value of polarization from using our current tilted optical Glauber model (labeled as ``CCNU") than from using SMASH and AMPT. This results from both the tilted geometry of the QGP fireball and the longitudinal flow gradient introduced in our current model. As discussed in the previous subsection, the tilted geometry also gives rise to the non-monotonic $p_\mathrm{T}$ dependence of the global polarization, which is absent in results from using the other two initialization models. When the tilt is strong, the magnitude of shear induced polarization increases rapidly with $p_{\rm{T}}$. On the other hand, this shear term from SMASH or AMPT initial condition only increases moderately in the given $p_{\rm{T}}$ region. Currently, it is hard to distinguish between the three initialization models based on the experimental data due to their large uncertainties. Future measurements with higher precision may help better constrain the initial condition in heavy-ion collisions.

%-------- Sec 3-4 ------
\subsection{Relation between global polarization and directed flow}
\label{sec:3-4}

\begin{figure}[tbp!]
\includegraphics[width=0.45\textwidth]{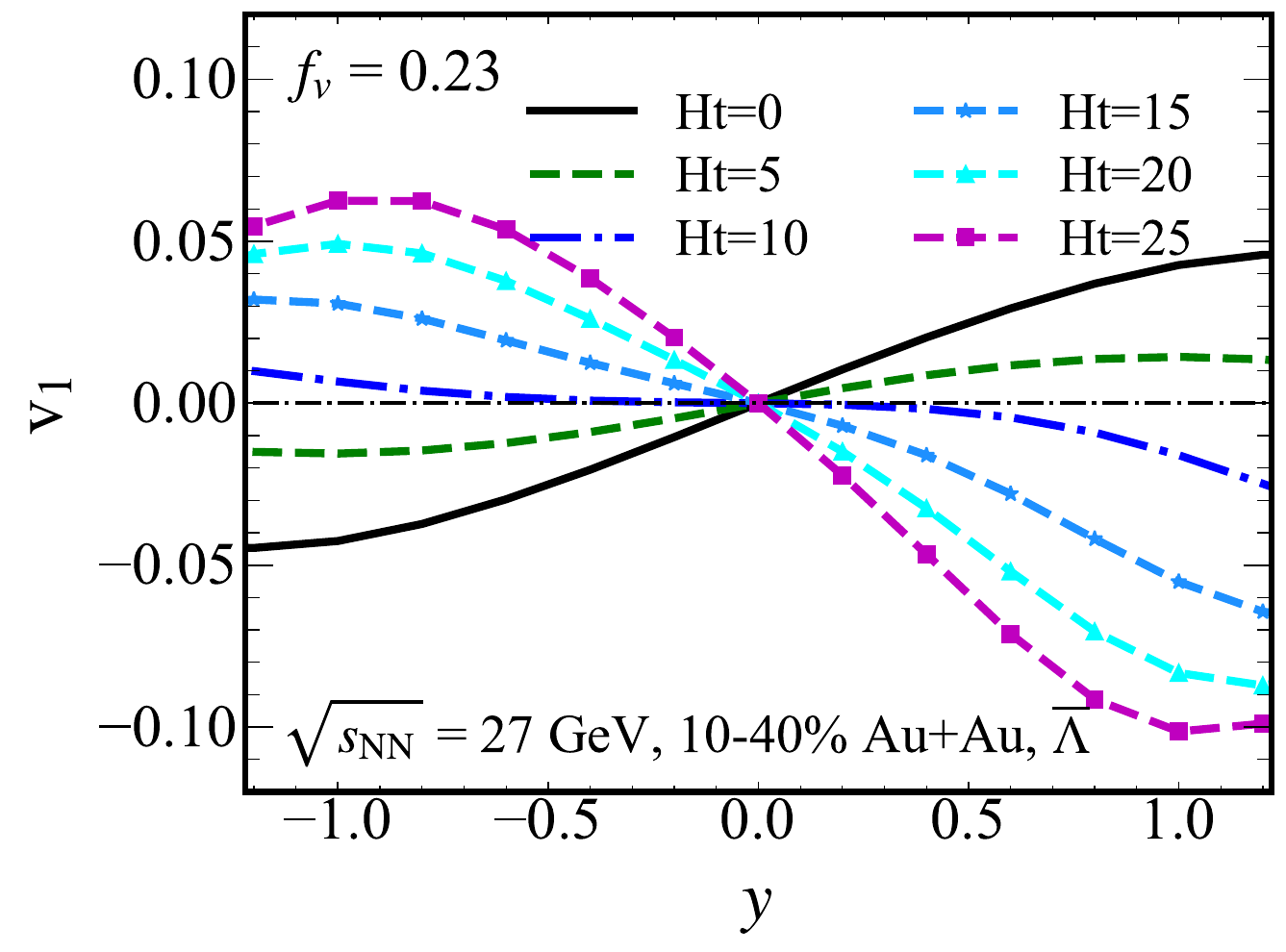}
\includegraphics[width=0.45\textwidth]{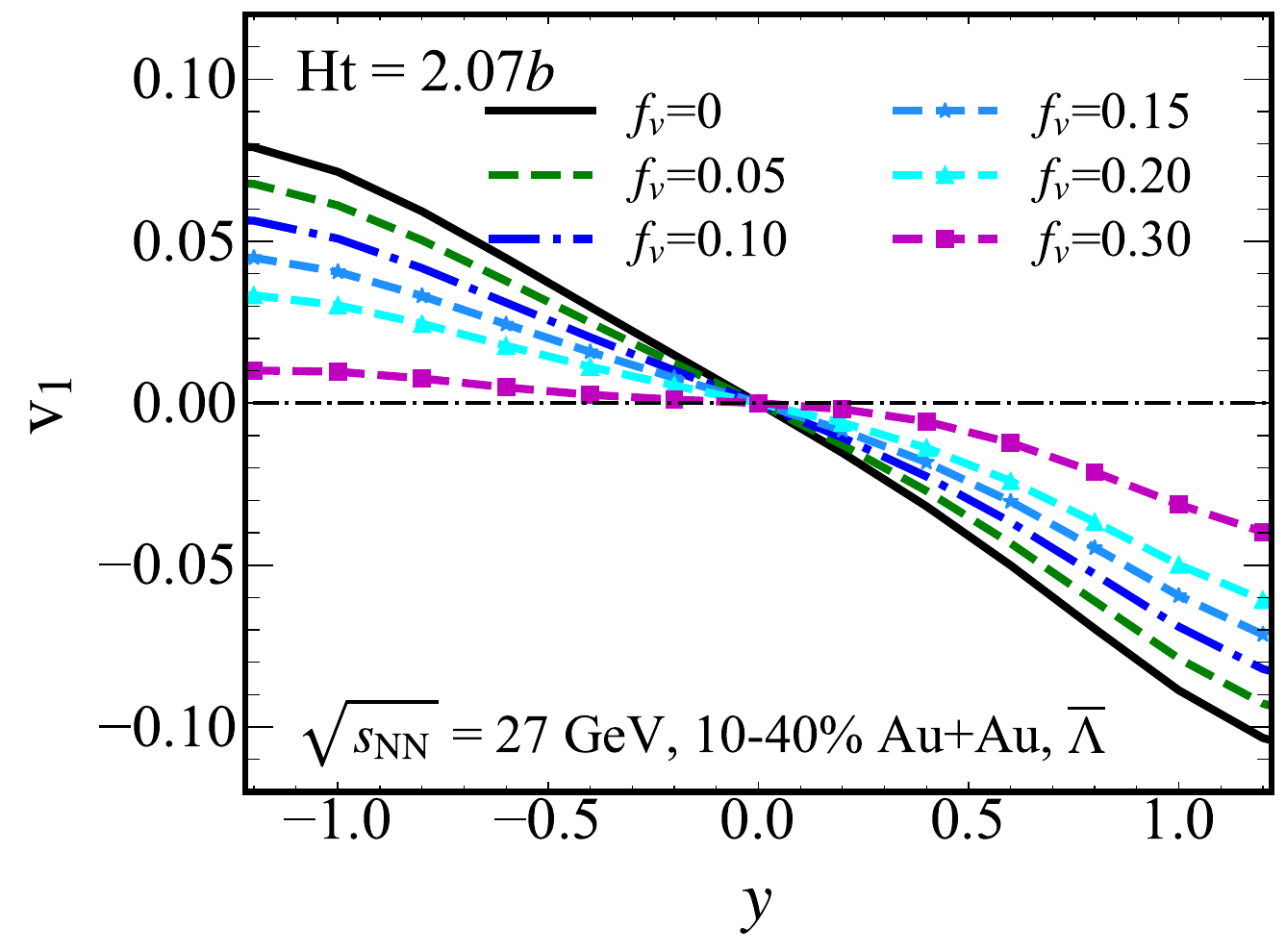}
\caption{(Color online) The directed flow of $\bar{\Lambda}$ as a function of rapidity, compared between different values of $H_\text{t}$ with $f_v$ fixed at 0.23 (upper panel), and between different values of $f_v$ with $H_\text{t}$ fixed at 14.8 (lower panel).}
\label{f:v1vsHf}
\end{figure}

\begin{figure}[tbp!]
\includegraphics[width=0.45\textwidth]{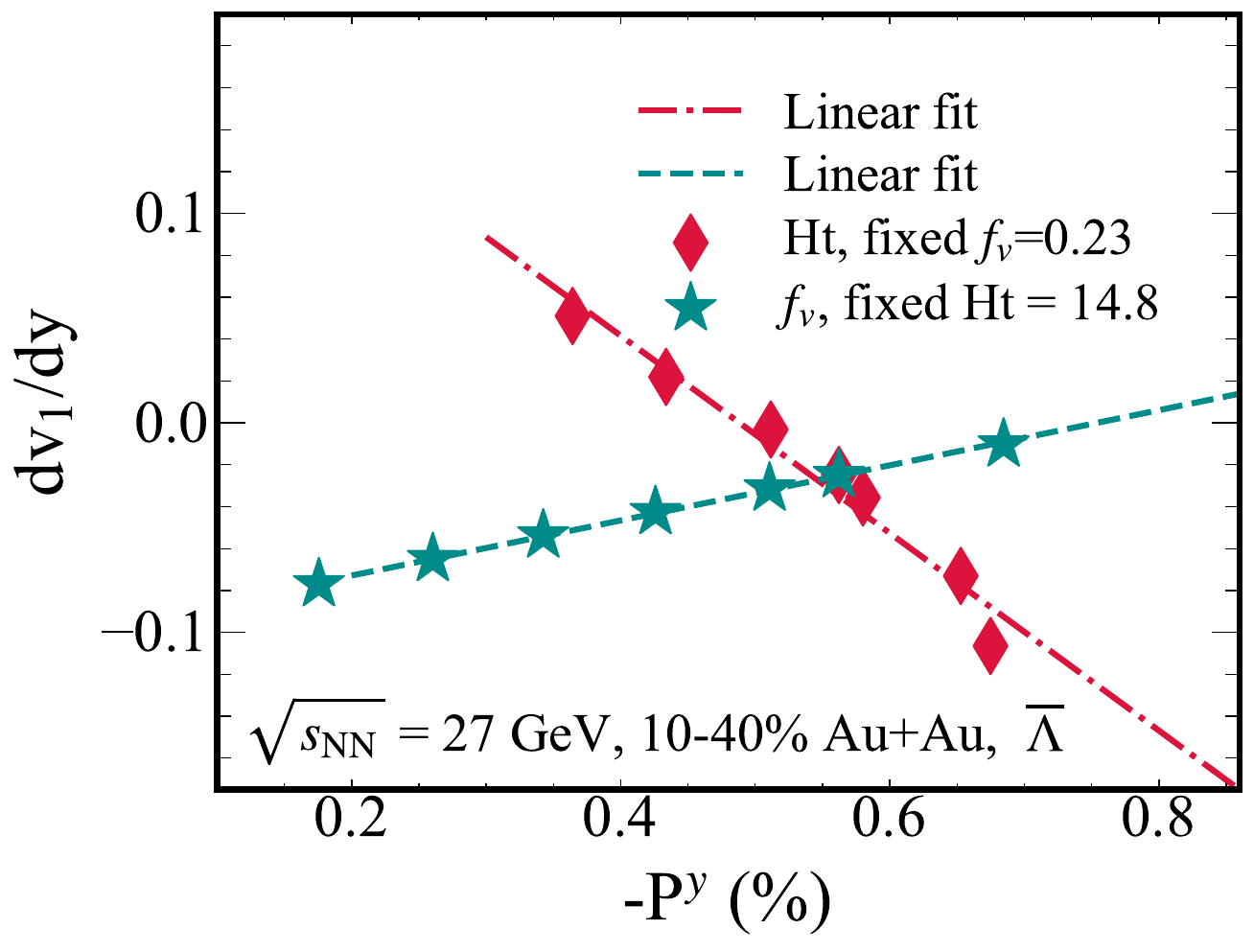}
\caption{(Color online) Relation between directed flow and global polarization of $\bar\Lambda$ with varying $H_\text{t}$ (red diamond symbols) or varying $f_v$ (green star symbols).}
\label{f:v1andpy}
\end{figure}

As seen in the previous two subsections, the value of hyperon polarization strongly depends on the initial condition of the QGP. Meanwhile, the initial geometry and flow field of the QGP also determine the collective flow coefficients of the final state hadrons. Therefore, one would naturally expect certain relation between these two observables in heavy-ion collisions, as already suggested by both experimental data~\cite{STAR:2017ckg,STAR:2019erd,STAR:2023nvo} and theoretical studies~\cite{Ivanov:2020wak,Ryu:2021lnx}. In this subsection, we will combine our analyses on the directed flow and global polarization of hyperons and explore how they are related to each other. 

Similar to Figs.~\ref{f:pol_ht}-\ref{f:pol_fy_total}, we first review the dependence of the hadron $v_1$ on the tilted geometry and the initial longitudinal flow profile in Fig.~\ref{f:v1vsHf} for 10-40\%  Au+Au collisions at $\sqrt{s_\text{NN}}=27$~GeV. Here we choose the $\bar\Lambda$ hyperon since the anisotropy of the anti-baryons is mainly driven by the energy distribution of the QGP rather than the baryon number density deposited by the projectile and target nuclei~\cite{Jiang:2023fad}. In the upper panel, we fix the $f_v=0.23$ parameter for the initial longitudinal flow and vary the $H_\text{t}$ parameter for the tilted deformation of the medium geometry. One observes that as $H_\text{t}$ increases from 0 to 25, the slope of directed flow with respect to rapidity ($dv_1/dy$) around mid-rapidity decreases from positive to negative values. On the other hand, when we fix $H_\text{t}=14.8$ (using $H_\text{t}=2.07b/\text{fm}$) for the medium geometry and vary $f_v$ for the longitudinal flow in the lower panel, one observes an increase in $dv_1/dy$ from negative values towards 0. These observations are consistent with our findings for anti-baryons in a prior work~\cite{Jiang:2023fad} on the directed flow coefficients of different hadron species at the BES energies. 
 
In Fig.~\ref{f:v1andpy}, we combine results of $dv_1/dy$ and $-P^y$ of $\bar{\Lambda}$ around mid-rapidity from our hydrodynamic calculation using different values of $H_\mathrm{t}$ and $f_v$. According to Figs.~\ref{f:pol_ht_total},~\ref{f:pol_fy_total} and~\ref{f:v1vsHf}, when $f_v=0.23$ is fixed, increasing $H_\text{t}$ increases $-P^y$ but decreases $dv_1/dy$. This leads to an almost linear decrease of the slope of the directed flow of $\bar{\Lambda}$ as its polarization increases, as shown by the red diamond symbols in Fig.~\ref{f:v1andpy}. Contrarily, when $H_\text{t}=14.8$ is fixed, increasing $f_v$ simultaneously increases $dv_1/dy$ and $-P^y$ of $\bar{\Lambda}$, resulting in an almost linear increase of the former with respect to the latter, as shown by the green star symbols. Therefore, as suggested in Ref.~\cite{Voloshin:2017kqp}, between  directed flow and global polarization, one may infer the value of one from the other. 

%-------- Sec 3-5 ------
\subsection{Local polarization of $\Lambda$ hyperons}
\label{sec:3-5}

\begin{figure}[tbp!]
\includegraphics[width=0.45\textwidth]{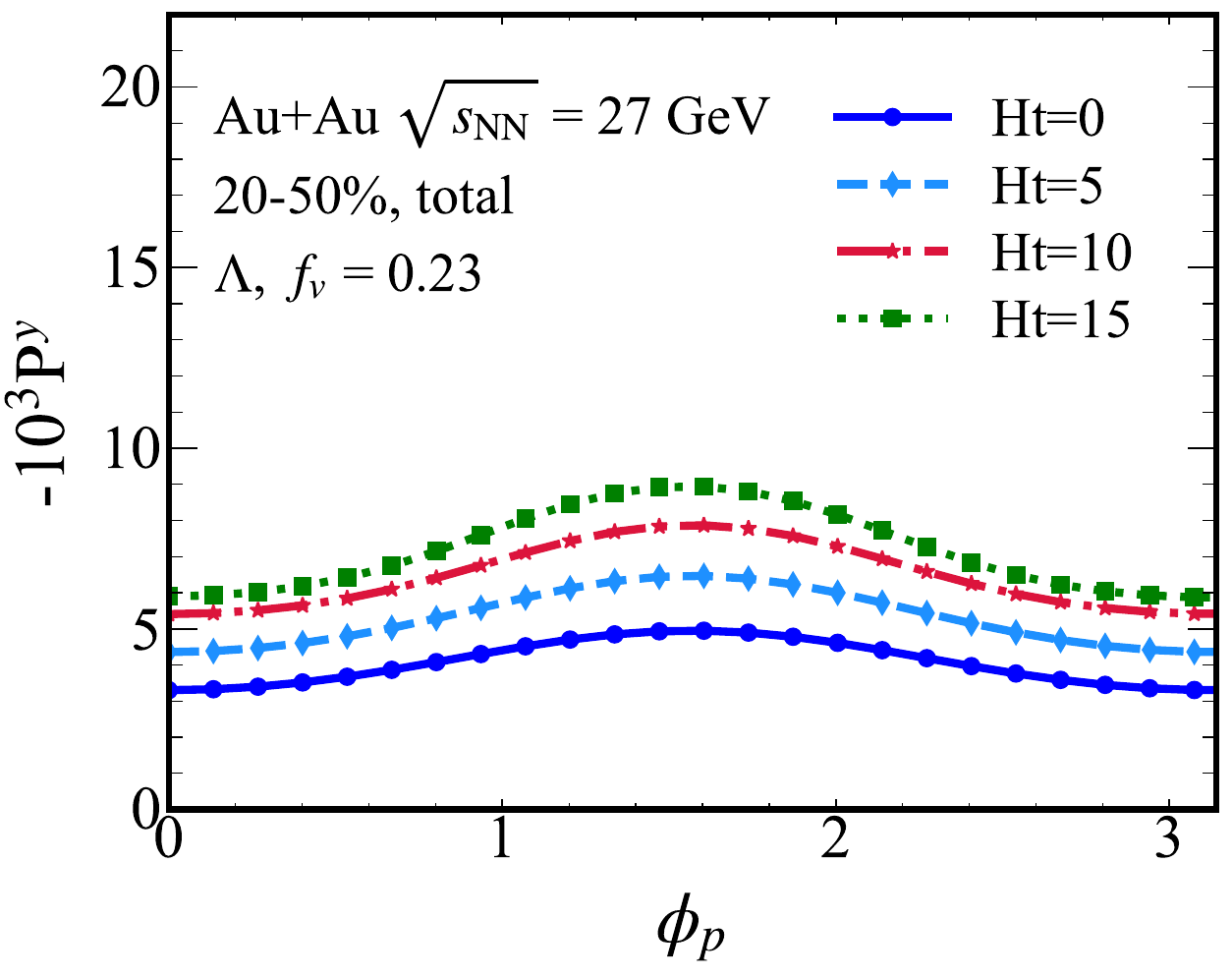}\\
\includegraphics[width=0.45\textwidth]{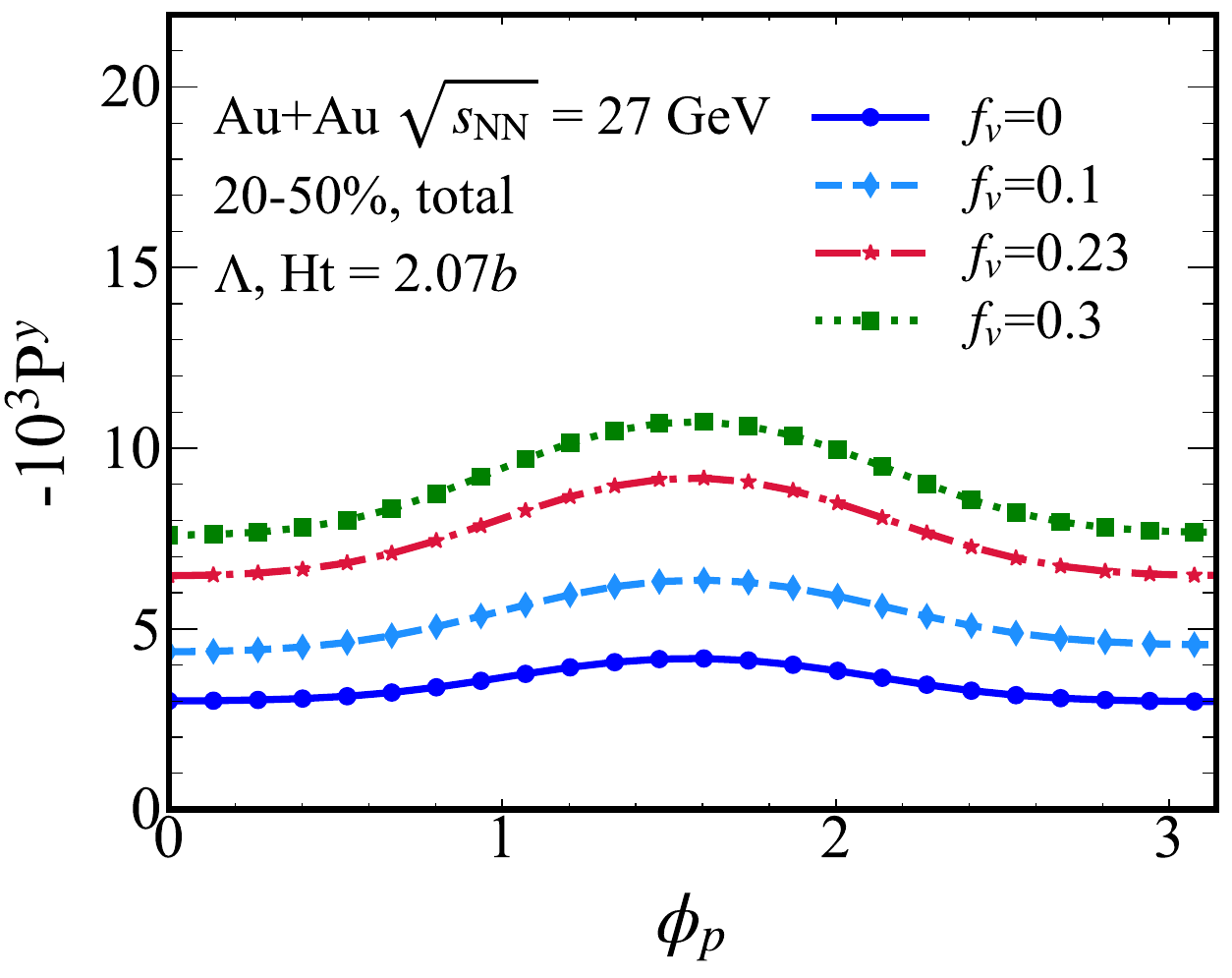}
\caption{(Color online) The local polarization along the $-\hat{y}$ direction of $\Lambda$ and $\bar{\Lambda}$ hyperons, $-P^{y}$, as a function of the azimuthal angle $\phi_{p}$ in 20-50\% Au+Au collisions at $\snn=27$ GeV.}
\label{f:local_py}
\end{figure}

\begin{figure}[tbp!]
\includegraphics[width=0.45\textwidth]{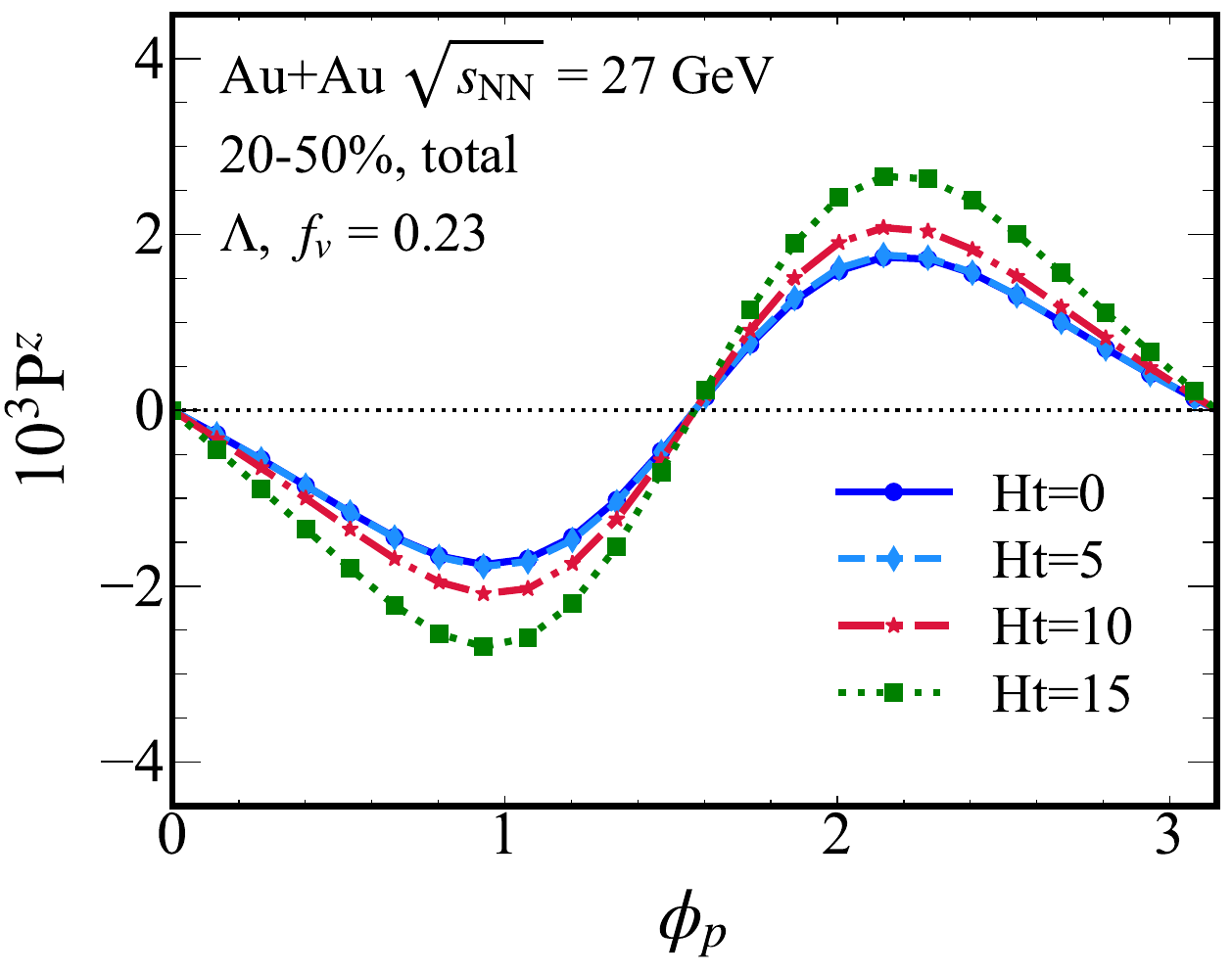} \\
\includegraphics[width=0.45\textwidth]{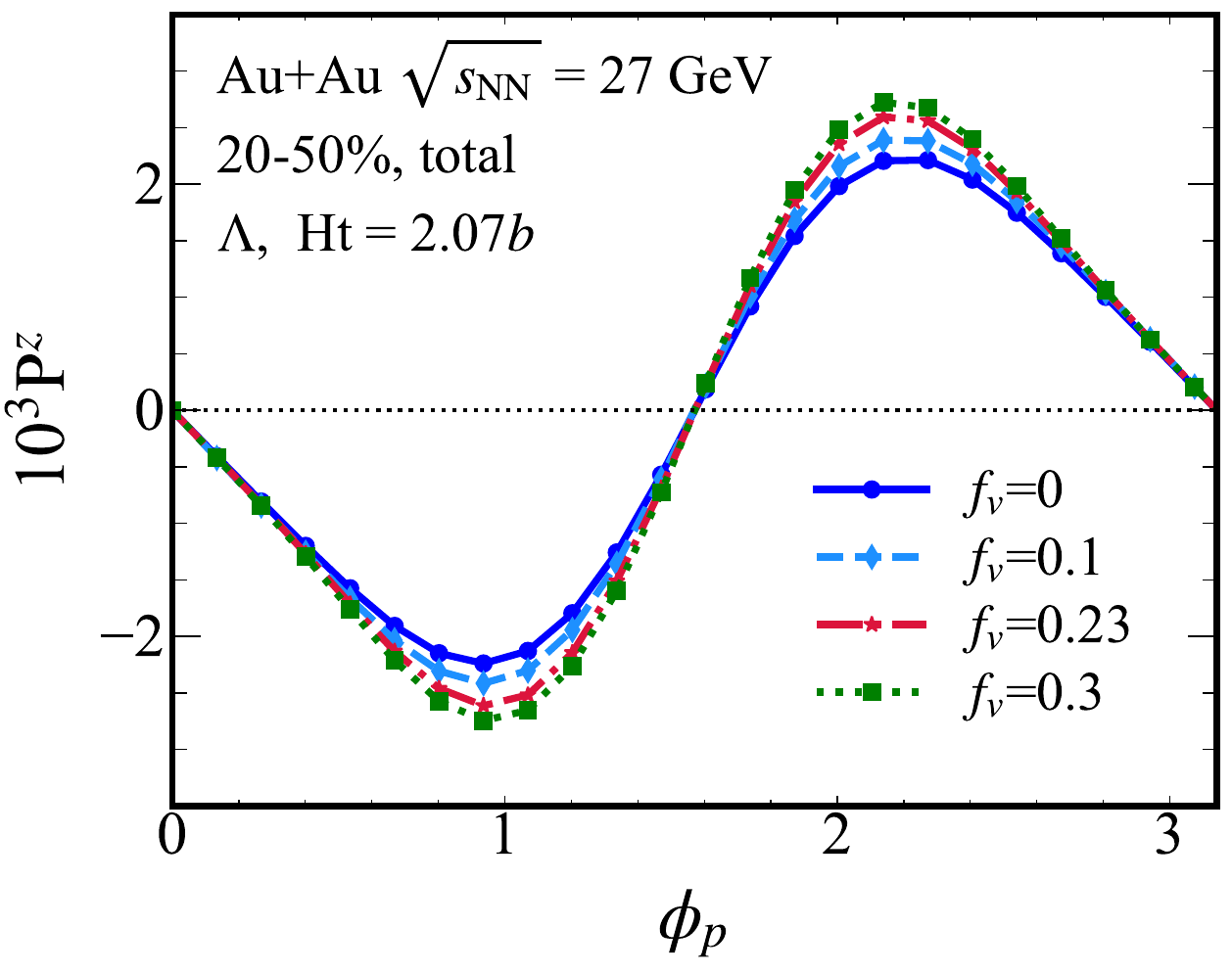}
\caption{(Color online) The local polarization along the beam direction of $\Lambda$ and $\bar{\Lambda}$ hyperons, $P^{z}$, as a function of the azimuthal angle $\phi_{p}$ in 20-50\% Au+Au collisions at $\snn=27$ GeV.}
\label{f:local_pz}
\end{figure}

In the end, we complete our study by presenting the local polarization of $\Lambda$ hyperons. 

Shown in Fig.~\ref{f:local_py} is the local polarization in the $-\hat{y}$ direction as a function of the azimuthal angle ($\phi_{p}$) in 20-50\% Au+Au collisions at $\snn=27$~GeV, compared between CLVisc hydrodynamic calculations using different $H_\text{t}$ (upper panel) and $f_v$ (lower panel) parameters. Consistent with our previous conclusions on the global polarization, enhancing the tilted deformation of the QGP or its initial longitudinal flow gradient also increases the local value of $-P^y$ at different $\phi_{p}$ between 0 and $\pi$. Similarly, increasing $H_\text{t}$ and $f_v$ also enhances the magnitude of local polarization in the $z$ direction ($|P^z|$), as shown in the upper and lower panels of Fig.~\ref{f:local_pz} respectively. Note that the cosine-like feature of $-P^y$ and the negative sine shape of $P^z$ with respect to $\phi_p$ are both opposite to observations in the experimental data~\cite{Niida:2018hfw,STAR:2019erd}. Although it has been proposed that contributions from the shear induced term and the spin Hall term help improve the theoretical description of local polarization towards the experimental observation~\cite{Becattini:2021iol,Fu:2021pok,Fu:2022myl}, after combining them with the dominating term of thermal vorticity, the discrepancies still exist in our current results.

%%%%%%%%%%%%%%%%%%%%%%% Sec - Conclusion

\section{Conclusions}
\label{v1section4}

We have studied the hyperon polarization and its relation with the directed flow of hadrons in Au+Au collisions at $\snn=27$~GeV. The CLVisc hydrodynamic simulation is coupled to a modified 3D Glauber initial condition that models a tilted QGP medium with an initial longitudinal velocity field. Using model parameters determined by the directed flow coefficient of different species of identified particles, our calculation provides a satisfactory description of the global polarization of $\Lambda(\bar{\Lambda})$ hyperons observed at the STAR experiment, as functions of centrality, transverse momentum and rapidity. We find that the thermal vorticity dominates the $p_\mathrm{T}$-integrated global and local polarization of hyperons, while the shear-induced polarization is important at high $p_\mathrm{T}$. Increasing the counterclockwise tilt of the QGP fireball with respect to the beam direction enhances the thermal vorticity contribution to the $\Lambda$ polarization at low $p_\text{T}$, while suppresses its contribution at high $p_\text{T}$. The opposite trend is found for the shear-induced contribution. Therefore, a non-monotonic dependence on $p_\mathrm{T}$ is found for the global polarization of $\Lambda$ with the presence of a tilted QGP profile. Effects of this tilted geometry on the fluid acceleration term and the baryonic spin Hall term are found small in our calculation. Depositing stronger initial longitudinal flow velocity into the QGP gives rise to a larger orbital angular momentum and therefore a larger thermal vorticity contribution to the $\Lambda$ polarization. However, effects of this initial velocity on the other three terms of polarization are found negligible. Compared to the same hydrodynamic simulation using SMASH or AMPT initial condition, our current calculation provides a larger value of $\Lambda$ polarization, indicating the sensitivity of global polarization to the initial geometry and the longitudinal flow velocity of the QGP. Furthermore, when the medium geometry is fixed, the slope of the $\Lambda$ hyperon $v_1(y)$ near mid-rapidity increases almost linearly with its global polarization as the initial longitudinal flow velocity is varied. On the contrary, a relation of linear decrease between these two quantities is observed when the initial flow is fixed while the tilt of the medium is varied. These imply the medium geometry and the longitudinal flow velocity are the common origin of polarization and directed flow, and therefore the combination of these two observables may provide a tight constraint on the initial condition of the QGP produced in non-central heavy-ion collisions. 
  
The framework presented in this work can be extended to studying the hyperon polarization at other beam energies at RHIC and LHC. However, apart from the medium geometry and longitudinal flow profile, other effects might be crucial for understanding the polarization phenomenology at lower collision energies. For instance, the electromagnetic field produced in energetic nuclear collisions can cause directional drift of charged quarks and thus affect the splitting of global polarization between $\Lambda$ and $\bar{\Lambda}$~\cite{Peng:2022cya,Guo:2019joy,Xu:2022hql}. 
The deformation of nuclear structure may also contribute to the polarization of hyperons~\cite{Hammelmann:2019vwd,Xi:2023isk,Ma:2022dbh,STAR:2023eck,Jing:2023zrh}. 
Finally, since all final-state observables are affected by the initial-state settings of the medium evolution, involving more observables, including the directed flow and global polarization, into the Bayesian analysis on model parameters would further improve our knowledge on the initial condition of the QGP.
These aspects will be explored in our upcoming efforts.

\begin{acknowledgements}
This work was supported by the National Natural Science Foundation of China (NSFC) under Grant Nos.~11935007, 12175122 and 2021-867, Guangdong Major Project of Basic and Applied Basic Research No.~2020B0301030008, the Natural Science Foundation of Hubei Province No.~2021CFB272, the Education Department of Hubei Province of China with Young Talents Project No.~Q20212703, the Open Foundation of Key Laboratory of Quark and Lepton Physics (MOE) No.~QLPL202104 and the Xiaogan Natural Science Foundation under Grant No.~XGKJ2021010016. 
\end{acknowledgements}

\bibliographystyle{unsrt}
\bibliography{clv3_v1ref}

\end{document}